\documentclass[]{aa}  
\usepackage{graphicx}
\usepackage{xcolor}
\usepackage{txfonts}
\usepackage{natbib}
\usepackage[modulo,switch]{lineno}
\bibpunct{(}{)}{;}{a}{}{,} % to follow the A&A style
\def\cd  {{$\mbox{c~d}^{-1}$}}

\def\msun {{\mathrm{M}_\odot}}
\def\sun {{_{\odot}}}
\def\rsun {{R\sun}}
\def\simless{\mathbin{\lower 3pt\hbox
     {$\rlap{\raise 5pt\hbox{$\char'074$}}\mathchar"7218$}}}   %< or of order
\def\simmore{\mathbin{\lower 3pt\hbox
     {$\rlap{\raise 5pt\hbox{$\char'076$}}\mathchar"7218$}}}   %> or of order

\newcommand{\ks}    {KS\,1947+300}

\newcommand{\sax}   {SAX\,J2103.5+4545}

\newcommand{\vcas}  {4U\,0115+63}

\newcommand{\bd}   {4U\,2206+54}

\newcommand{\ha}   {H$\alpha$}

\begin{document}
   \title{Fast time optical variability in Be/X-ray binaries. \\
   Pulsation and rotation}

   \author{
P. Reig\inst{1,2} \and
J. Fabregat\inst{3}
}

   \institute{
%1
Institute of Astrophysics, Foundation for Research and Technology, 71110 Heraklion, Crete, Greece
\and
%2
University of Crete, Physics Department \& Institute of
Theoretical \& Computational Physics, 70013 Heraklion, Crete, Greece
\and
%3
Observatorio Astron\'omico de la Universidad de
Valencia, Calle Catedr\'atico Agust\'\i n Escardino 7, 46980 Paterna, Valencia, Spain
}

%Institute for Astronomy (IfA), University of Vienna,
%              T\"urkenschanzstrasse 17, A-1180 Vienna\\
%              \email{wuchterl@amok.ast.univie.ac.at}
%         \and
%             University of Alexandria, Department of Geography, ...\\
%             \email{c.ptolemy@hipparch.uheaven.space}
%             \thanks{The university of heaven temporarily does not
%                     accept e-mails}
%             }

   \date{Received ; accepted }

% \abstract{}{}{}{}{} 
% 5 {} token are mandatory
 
\abstract 
{
Classical Be stars, regardless of spectral subtype, display multiperiodic light
modulations in the frequency range 0.1-12 \cd, when observed with
high-cadence and long duration.  This behaviour is attributed to non-radial
pulsations and/or rotation of the Be star. A similar study for the optical
counterparts to Be/X-ray  binaries is yet to be carried out.
}
{The main goal of this work is to investigate the fast photometric variability of
the optical counterparts to Be/X-ray binaries and compare the general
patterns of such variability with the Galactic population of classical Be
stars. 
}
{The main core of our analysis is based on space-based observations performed by TESS.  We analyzed 21 sources with TESS.  High-cadence photometry with two ground-based telescopes was also performed for 15 sources.  The TESS light 
curves were created from the full-frame images using the {\tt Lightkurve} package. 
The ground-based light curves were obtained  through differential photometry
between the target and a number of non-variable  stars in the same field of
view. Standard Fourier analysis and least-squares fitting methods were employed 
in the frequency analysis. } 
{ All sources exhibit intra-night light variations with intensity variations of 0.01-0.06 mag in the ground-based observations and up to 5\% in flux in TESS observations. This
variability manifests itself as multi-periodic signals in the frequency range 0.2--12
\cd. We find that  the patterns of variability of the Be stars in Be/X-ray
binaries  agree with that of classical early-type Be stars in terms of the
general shape of the periodograms. Based on the general shape and number of
peaks in the periodograms, Be/X-ray binaries can be classified into different types. The most
common case is the presence of groups of closely-spaced frequencies (67)\%,
followed by sources that exhibit isolated signals (18\%). The
remaining type of sources displays frequency spectra characterized by a mixed pattern of stochastic
variability and high-frequency peaks.  }
{This study reveals that short-term optical photometric variability is a very common, if not 
ubiquitous, feature intrinsic to the Be optical companions in Be/X-ray binaries. This variability is
mainly attributed to pulsations that originate in the stellar interior.  }


   \maketitle
   
%\linenumbers
%
%________________________________________________________________

\section{Introduction}

%NRPs appear as the most likely  mechanism that can explain the short-term
%periodicities with frequencies above $\sim$2 \cd detected in the light  curves
%of their optical counterparts.

%The term "Be phenomenon" refers to the episodic ejection of mass from the
%atmosphere of early-type  stars that lead to the formation of a flattened disk
%around their equator \citep{porter03,rivinius13a}. The evidence for the
%existence of such disks stems from the presence of emission lines, infrared excess, and
%intrinsic polarization. The stars that show the Be phenomenon are known as Be
%stars, although late O-type star may also host circumstellar disks.
%The lines most affected by emission are those of the Balmer series, but in some
%cases the profile of \ion{He}{I} and \ion{Fe}{II} lines also appear  distorted
%as a result of partial or total fill-in emission \citep{hanuschik96b}.  

High-mass X-ray binaries (HMXB) divide into supergiant X-ray binaries (SGXB) and
Be/X-ray binaries (BeXBs) according to the luminosity of the optical
counterpart. In a SGXB a neutron star orbits around an evolved (luminosity class
I or II) star  while in BeXBs the optical companion is a giant or main-sequence (luminosity class III--V) star. In Be stars, episodic ejection of mass from their atmospheres results in the formation of a flattened disk around their equator
\citep{porter03,rivinius13a}. The evidence for the existence of such disks stems
from the presence of emission lines, infrared excess, and intrinsic
polarization.  The ultimate mechanism that injects material at the Keplerian
orbital speed into the base of the disk is still unknown. Centrifugal
acceleration is a promising mass loss mechanism because Be stars rotate, on
average, at larger speeds than normal B stars. Although there is now general
consensus that Be stars are rapidly-rotating stars, there is no agreement on how
close to critical velocity they rotate. Rotation rates (rotation velocity
normalized by the star's critical rotation speed) lie in the range 70-80\%
\citep{porter03,rivinius13a}, 90\% \citep{fremat05} or even higher \citep{townsend04}. The
rotation rate is a key ingredient in the understanding of the formation of the
equatorial disk. The higher the rotation rate the easier to lift up gas and
launch material into a ballistic orbit. But if Be stars rotate below critical
velocity, then the launching mechanism still has to be identified. Non-radial
pulsations (NRP), probably in combination with small-scale magnetic fields have
been suggested as a possible mechanism to explain mass-loss and disk-formation
in hot stars \citep[][]{baade92,cranmer09,rivinius13a,baade18b}. 
%Critics of this model claim that because the additional velocity provided by NRP is small, this mechanism would imply that all Be stars should have rotational velocities in
%excess 90\% of the critical rotational velocity. 
An alternative mechanism that
may eject material from the photosphere to form the disk would be magnetic
reconnection associated with localized magnetic fields generated by convection
\citep{balona03,balona20,balona21}. 

Rotation and pulsations manifest as multi-frequency photometric variability, that is, as brightness variations \citep{baade16,rivinius16,semaan18,balona20,balona21} and as spectroscopic variability, e.g. as line profile variations \citep{baade84,rivinius98,rivinius03,balona03,zima06} with
typical periods in the range 0.1 to 2 days. Therefore, the study of the
short-term variability in Be stars may provide the key to distinguish between
competitive models.

The optical emission in a BeXB comes from the circumstellar disk and from the
early-type star. A third source of optical emission would be an accretion disk
around the neutron star. However, the optical flux from the accretion disk is
several orders of magnitude lower than the optical emission from the Be star.
While there are many studies on the evolution of the decretion disk, little has
been done to understand the physical properties of the underlying Be star other
than the determination of the spectral type. It is generally assumed that the Be
star in BeXBs has the same properties as a classical Be star. In other words,
for the same spectral type, the luminosity, mass, and radius are similar.
The traditional view has been that due to the large mass ratio and wide
orbital separation between the Be star and the neutron star,  the neutron star
is not expected to exert a significant effect on processes related to the
surface of the optical companion. Therefore, the same phenomenology in the
short-term variability is expected between classical Be stars and
BeXBs. We know now that this is not true for the long-term variability that it
is associated with the disk.  Disks in BeXBs are denser and smaller than in
classical Be stars \citep{reig97a,zamanov01,reig11,zamanov13,reig16}. The reason is that
disks in BeXBs are truncated by the tidal torque  exerted by the neutron star. The distance at which the circumstellar disk is truncated depends mainly on the orbital parameters and the viscosity \citep{okazaki01,okazaki02}. In addition, in a truncated disk, the large amount of material that has to be accreted to produce the giant X-ray outbursts can only be achieved if the disk is misaligned and warped \citep{martin11,okazaki13,martin14a}. 
Since truncation results in the
limitation of the size of the equatorial disk of the Be star in a BeXB,   one may wonder 
whether the optical companions in BeXBs may also show different phenomenology
with regard to processes associated with the star itself. Although there is growing evidence that classical Be stars harbor low-mass companions \citep{gies00,klement22} that can also lead to the truncation of the disk \citep{panoglou16,cyr17,klement17,klement19}, the tidal interaction is expected to be less significant and accretion is not expected to be relevant as in BeXBs.  The viscous decretion disk model explains quite satisfactorily  most of the Be star's phenomenology \citep{haubois12,haubois14,draper14}.

Short-term photometric periodic variability is commonly observed in classical Be stars \citep[][and references therein]{labadie17,bernhard18}. Be stars that have been observed with
high-cadence and long-duration, especially from space-based observatories, have
shown multi-periodic variability that in most cases have been attributed to NRP
\citep{gutierrez-soto07b,huat09,neiner09,emilio10,kurtz15,rivinius16,baade16,semaan18,baade18b,balona20,labadie22}. 
In contrast, only two Galactic BeXBs have been reported to display periodicities
of the order of a few hours in their photometric light curves: GRO\,J2058+42
\citep{kiziloglu07b}, and LS I +61 235 \citep{sarty09}. A tentative detection of
a 2.6 \cd\ periodicity has also been reported in \bd\ \citep{bugno09}. This
scarcity of results is not due to the lack of detections but to the scarcity of
observations.  To be able to perform meaningful comparative studies between the
rotational and pulsational properties of Be stars in different types of systems
and environments, the sample of Milky Way Be stars in X-ray binaries must be
increased. To remedy this, we set up a new project to investigate the fast time
variability of BeXBs visible from the Northern hemisphere from ground-based
telescopes. 
 The launch of the Transiting Exoplanet Survey Satellite (TESS) added a new dimension to the project and gave us the opportunity, not only to expand our target list to systems visible from the Southern hemisphere but also to expand the frequency space. Thus, our analysis
is mainly based on the TESS light curves. We  used our ground-based photometry to
complement the space data, especially for the fainter sources not observed by
TESS. TESS observations of classical Be stars have been presented in \citet{labadie21,labadie22} and \citet{balona20,balona21}.

%-------------------------------------------------------------
\begin{table*}
\caption{Target list. The fourth column indicates the type of variability of the periodograms as defined in Sect.~\ref{periodograms} : $g$: frequency groups, $i$: isolated frequencies, $s$: stochastic variability, and $h$: high-frequency signals (applicable to TESS data only).}             
\label{target}      
\centering          
\begin{tabular}{l@{~~}l@{~~}c@{~~}c@{~~}c@{~~}c@{~~}c@{~~}c@{~~}c@{~~}l@{~~}}
\hline\hline
Source		&Spectral   &$V$-band   &Type       &P$_{\rm spin}$ &P$_{\rm orb}$ &$e$ 	&Distance (kpc)		&$v\sin i$	&References\\
name		&type		&(mag)      &var.       &(s)   	&(days)		&	&Gaia EDR3$^\ddagger$	&(km s$^{-1}$)  	&\\
\hline
IGR\,J00370+6122&BN0.7Ib    &9.6	   &s       &346	&15.66		&0.5	&$3.4\pm0.2$		&$135\pm7$	&[1a], [1b] \\
2S\,0114+65	&B1Ia           &11.0	   &s       	&9720	&11.6		&0.18	&$4.9^{+0.3}_{-0.2}$	&$96\pm20$	&[2a], [2b]  \\
4U\,0115+63	&B0.2Ve         &15.4	   &g       	&3.6	&24.3		&0.34	&$5.8^{+0.8}_{-0.5}$	&$300\pm50$   	&[3a], [3b] \\
IGR\,J01363+6610&B1IV-Ve    &13.3	   &g       &--	&--		&--	&$5.8\pm0.4$		&$160\pm20$	&[4a], [4b], [4c]\\
RX\,J0146.9+6121$^\dag$&B1Ve&11.3	   &--       	&1400	&303?		&--	&$2.75^{+0.16}_{-0.14}$	&$200\pm30$   	&[5a], [5b] 	 \\
%IGR\,J01583+6713&B2IVe    &		   &       &--	&--		&--	&$6.1^{+0.5}_{-0.4}$	&$330\pm30$	&[5a], [5b], [5c] \\
RX\,J0240.4+6112&B0Ve       &10.8	   &g       	&--	&26.5		&0.54	&$2.5\pm0.7$		&$350\pm8$	&[6a], [6b]	\\
SWIFT\,J0243.6+6124&O9.5Ve    &12.8	   &g         &9.9	&27.6		&0.10	&$5.2\pm0.3$		&$210\pm20$   	&[7a], [7b]  	 \\
V\,0332+53	&O8--9Ve        &15.4	   &g,h       &4.4	&33.8		&0.37	&$5.6^{+0.7}_{-0.5}$	&$150\pm30$	&[8a], [8b]	\\
X\,Per		&B0Ve           &6.1	   &g,h       	&838	&250		&0.11	&$0.60^{+0.02}_{-0.01}$	&$215\pm10$	&[9a], [9b]	\\
RX\,J0440.9+4431&B0.2Ve    &10.7	   &g,h       	&202.5	&150		&--	&$2.44^{+0.06}_{-0.08}$	&$235\pm15$   	&[10a], [10b]  	 \\
1A\,0535+262	&O9.7IIIe    &9.2      &g,h      	&105	&111		&0.47	&$1.79^{+0.08}_{-0.07}$	&$225\pm10$   	&[11a], [11b] 	 \\
IGR\,J06074+2205&B0.5Ve    &12.2	   &i,h       	&373.2	&--		&--	&$6.0\pm0.6$		&$260\pm20$   	&[12]  	 \\
MXB\,0656--072	&O9.5Ve    &12.3	   &i       	&160.4	&5101.2		&0.4?	&$5.7\pm0.5$		&--		&[13a], [13b]	\\
RX\,J0812.4--3114&B0.2IVe    &12.7	   &g,h       &31.9	&80		&--	&$6.7^{+0.5}_{-0.4}$	&$240\pm20$	&[14a], [14b]	\\
Vela X--1	&B0.5Ib         &6.9	   &s       	&283	&8.96		&0.09	&$2.00\pm0.06$		&$116\pm6$	&[15]	\\
GRO\,J1008-57	&B1Ve       &15.3	   &g,h       	&93.5	&249.5		&0.68	&$3.54\pm0.15$		&--		&[16a], [16b]	\\
RX\,J1037.5--5647&B0III-Ve    &11.5   &s       	&860	&--		&--	&$5.0^{+0.4}_{-0.3}$	&--		&[17a], [17b]	\\
1A\,1118--616	&O9.5IV-Ve    &12.1   &g,h       &406.5	&24		&$<0.1$	&$2.90^{+0.09}_{-0.08}$	&$270\pm25$	&[18a], [18b]	\\
4U\,1145--61	&B0.2IIIe    &9.0	   &g,h       &292	&186.5		&$>0.5$	&$2.06^{+0.08}_{-0.09}$	&$280\pm30$	&[19a], [19b]	\\
GX\,304--1	&B0.7Ve         &14.4	   &i,h       	&272	&132.5		&0.46	&$1.90\pm0.05$		&$330\pm50$	&[20a], [20b]	\\
KS\,1947+300	&B0Ve       &14.5	   &s       	&18.7	&40.4		&0.03	&$15.1^{+3.2}_{-2.6}$  	&$310\pm40$ 	&[21a], [21b]  	 \\
Swift\,J2000.6+3210$^\dag$&B0-2Ve&16.2 &--       &890	&--		&--	&$8.4^{+1.8}_{-1.2}$	&-- 		&[22] 	 \\
GRO\,J2058+42	&O9.5-B0IV-Ve    &14.9   &i,h       	&192	&110		&--	&$8.9^{+1.1}_{-0.9}$	&$250\pm50$ 	&[23a], [23b]   \\
SAX\,J2103.5+4545$^\dag$&B0Ve    &13.9   &--       		&358	&12.7		&0.40	&$6.2^{+0.6}_{-0.5}$	&$240\pm20$ 	&[24a], [24b]  \\
IGR\,J21343+4738$^\dag$&B1IVe    &14.1    &--       		&320	&--		&--	&$8.3^{+0.9}_{-0.8}$    &$365\pm15$ 	&[25]    \\
Cep\,X--4$^\dag$	&B1-2Ve    &14.3   &--       		&66.3	&24.8?		&--	&$7.5^{+0.6}_{-0.5}$	&$460\pm50$ 	&[26]    \\
4U\,2206+54	&O9.5Ve         &9.8	   &s,h       	&5550	&19.2		&0.15	&$3.1^{+0.1}_{-0.1}$	&$315\pm70$ 	&[27]    \\
SAX\,J2239.3+6116&B0Ve       &14.4	   &g,h       	&1247	&263		&--	&$7.4^{+0.9}_{-0.7}$	&$195\pm20$ 	&[28]    \\
MWC\,656	&B1.5IIIe       &8.7	   &g       &--	&60.4		&0.10	&$2.0\pm0.1$		&$313\pm3$  	&[29]	  \\
\hline\hline
\end{tabular}
\tablefoot{$^\dag$  Ground-based data only; $^\ddagger$  \citet{bailer-jones21}. Gaia distances are model dependent and may differ substantially in the various Data Releases (DR). Here we quote the distance obtained from EDR3 using https://dc.zah.uni-heidelberg.de/gedr3dist/q/cone/form. See also \citet{arnason21}  ;  
References: 
[1a] \citet{gonzalez-galan14}; [1b] \citet{grunhut14};
[2a] \citet{crampton85}; [2b] \cite{reig96};
[3a] \citet{negueruela01a}; [3b] \citet{raichur10}; 
[4a] \citet{reig05a}; [4b] \citet{tomsick11}; [4c] $v\sin i$ (Reig, Priv. Comm.);
[5a] \citet{reig97b}; [5b] \citet{sarty09}; 
%%[5a] \citet{wang10}; [5b] \citet{kaur08}; [5c] (Reig, Priv. Comm.)
[6a] \citet{aragona09}; [6b] \citet{zamanov13}; 
[7a] \citet{wilson18}; [7b] \citep{reig20}; 
[8a] \citet{negueruela99}; [8b] \citet{doroshenko16}; 
[9a] \citet{lyubimkov97}; [9b] \citet{delgado01}; 
[10a] \citet{reig05b}, [10b] \citet{ferrigno13};
[11a] \citet{haigh04}; [11b] \citet{grundstrom07b}; 
[12] \citet{reig10b}; 
[13a] \citet{yan12a}; [13b] \citet{nespoli12};
[14a] \citet{reig01}; [14b] \citet{corbet00}; 
[15] \citet{vankerkwijk95}; 
[16a] \citet{kuhnel13}; [16b] \citet{coe07}; 
[17a] \citet{reig99}; [17b] \citet{motch97}; 
[18a] \citet{motch88}; [18b] \citet{staubert11}; 
[19a] \citet{janot82}; [19b] \citet{alfonso17}; 
[20a] \citet{parkes80}; [20b] \citet{malacaria17};
[21a] \citet{galloway04}; [21b] $v\sin i$ (Reig, Priv. Comm.); 
[22] \citet{masetti08}; 
[23a] \citep{wilson98}; [23b] \citep{kiziloglu07b};
[24a] \citet{baykal00};  [24b] \citet{reig04};
[25] \citet{reig14a}; 
[26] \citet{bonnet98}; 
[27a] \citet{corbet07}; [27b] \citet{blay06}; [27c] \citet{reig09};
[28] \citet{reig17};
[29] \citep{zamanov21}
}
\end{table*}
%-------------------------------------------------------------                                      

\section{Observations}

We extracted light curves for all the BeXBs observed by TESS. These are 25 sources, which represent about 80\% of the confirmed (i.e. with well constrained optical counterparts) Galactic BeXBs \citep{raguzova05}\footnote{We used the extended and updated version of this catalog available on-line: http://xray.sai.msu.ru/~raguzova/BeXcat}.  However, owing to the large pixel size of the TESS cameras (21 arcsec), 4 out of 25 sources were affected by nearby bright objects that prevented us from obtaining a clean light curve.  These are: RX\,J0146.9+6121, SAX\,J2103.5+4545, IGR\,J21343+4738, and Cep\,X--4. High-cadence photometry with two ground-based telescopes was also performed for 15 sources, including the four sources without clean TESS light curves. 
In order to compare the pulsational properties of BeXBs and SGXBs we
also analyzed two prototypical examples of SGXBs, namely, Vela X--1 and
2S\,0114+65, and the less luminous SGXB IGR\,J00370+6122. Table~\ref{target}
gives the list of sources analyzed in this work together with some astrophysical
information. 

%Table~\ref{logobs} in the appendix shows the log of the ground-based observations. 

\subsection{Transiting Exoplanet Survey Satellite (TESS)} 

TESS is optimized for the detection of exoplanets (super Earths) around nearby,
bright stars  \citep{ricker15}. Owing to its observational strategy it covers the entire
sky in periods of two years.   TESS observes the sky in sectors of $24^{\circ}
\times 96^{\circ}$.  Each sector is observed for two orbits of the satellite
around the Earth, or about 27 days on average. TESS has four identical cameras
equipped with custom f/1.4 lenses, providing each camera with a $24^{\circ}
\times 24^{\circ}$ field of view. The cameras have an effective aperture size of
10 cm in diameter. TESS uses a red-optical band-pass covering the wavelength range
from about 600 to 1000 nm, with maximum throughput between  700 and 900 nm.

%The Data Handling Unit (DHU) on board TESS provides two basic data
%products: two-minute stacked subarrays and 30-minute Full frame Images (FFIs). 
%During nominal operations, the four CCD cameras are clocked continuously, with
%an integration time of two seconds. The DHU uses data from 60 consecutive
%two-second integrations to create two-minute postage stamps and data from 900
%consecutive two-second images to create 30-minute FFIs for all cameras.

\subsection{Skinakas observatory}

The Skinakas observatory\footnote{https://skinakas.physics.uoc.gr/en/} (SKO) is located on the Ida mountain in central Crete
(Greece) at an altitude of 1750 m. The observations reported in this work  were
made with the 1.3m modified Ritchey-Chr\'etien telescope. Two ANDOR CCD
back illuminated cameras were employed. Before 2018, we used a DZ436. From
2018, an IKON-L 926. Both have an array of 2048 $\times$
2048 13.5 $\mu$m pixel size (corresponding to 0.28 arcsec on sky), hence
providing a field of view of 9.5 square arcmin.  Appendix~\ref{ground} gives the log of the observations from the Skinakas observatory.

\subsection{Aras de los Olmos observatory}

The Aras de los Olmos observatory\footnote{https://www.arasdelosolmos.es/observatorios/} (OAO) is located in la Muela de Santa
Catalina, near the Aras de los Olmos town (Valencia, Spain), at an altitude
of 1280 m. Observations were performed with the 0.5-m telescope equipped with a
Finger Lakes Instruments ProLine PL16801 thermoelectrically cooled CCD camera.
The 4096 $\times$ 4096 array with a 2 $\times$ 2 binnig gives a plate scale of
1.08$\arcsec$/pixel and a field of view of $36.9$ square arcmin.
Appendix~\ref{ground} gives the log of the observations from the Aras de los Olmos observatory. 

\section{Data analysis}

\subsection{TESS light curves}

The Full Frame Images (FFIs) are the basic data product of the TESS
mission. A single FFI is the full set of all science and collateral pixels
across all CCDs of a given camera. FFIs were taken every 30 minutes during
science operations in the primary mission (July 2018-July 2020) and with a
cadence of 10 minutes during the first extension of the mission\footnote{TESS data products, https://heasarc.gsfc.nasa.gov/docs/tess/data-products.html } (scheduled to end in September 2022). We created our own light curves from the FFIs by
using the {\tt Lightkurve} \citep{lightkurve} and {\tt TESScut} \citep{brasseur19} packages to download a Target Pixel File (TPF) with a 25 $\times$ 25 pixel image (8.75 $\times$ 8.75 arcmin, being 21 arcsec the projected pixel size), centered on the target for every available TESS sector. These images were carefully inspected and different aperture masks were chosen for the different targets as a function of their brightness and the presence of nearby brighter stars.  We removed unwanted events by
filtering out observations with a {\tt QUALITY} flag different from zero. In
practice, these events were mainly outliers.

%-------------------------FIG 1------------------------------------
   \begin{figure*}
   \centering
    \begin{tabular}{cc}   
     \includegraphics[width=8cm]{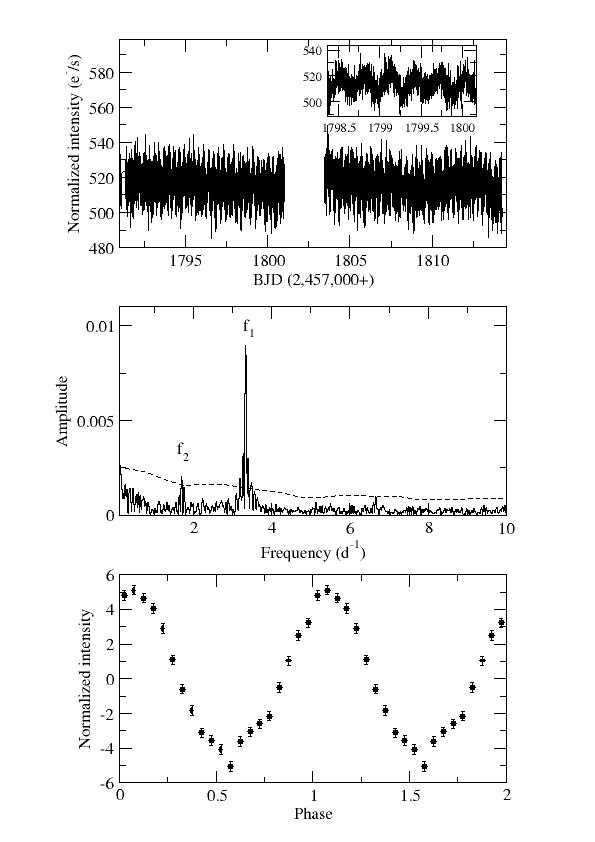} & 
     \includegraphics[width=8cm]{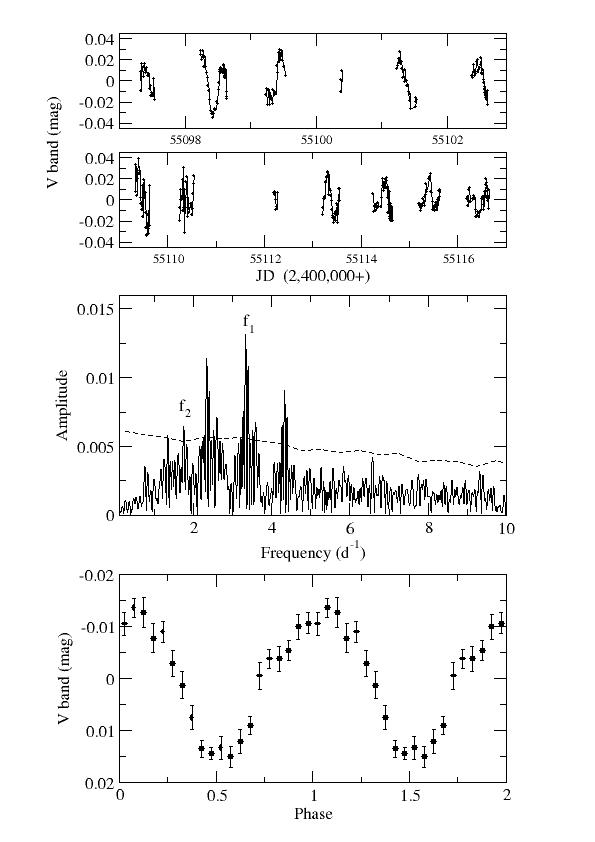} \\
    \end{tabular}
     \caption{Analysis of 4U\,0115+53 performed on TESS (left) and
     Skinakas (right) observatories.  {\em Top panels}: The light curves where 
      intra-night variability is clearly detected. {\em Middle panels}:
      Periodograms. 
      The dashed lines represent four times the average noise level
      of the final prewhitenning periodogram.  The frequencies on both sides of
      $f_1$ on the right panel are aliases at $f_1\pm1$ \cd and disappear after
      prewhitenning with $f_1$.
      {\em Bottom panels}: Phase diagrams folded with the  3.3 \cd frequency.
      }
         \label{4u0115}
   \end{figure*}

%_____________________________________________________________

\subsection{Ground-based light curves}

Reduction of the ground-based telescope data were carried out in the standard
way using the IRAF tools for aperture photometry \footnote{A User's Guide to CCD Reductions with IRAF, Philip Massey, February 1997. https://iraf.net/irafdocs}  \citep{tody86}. First, all CCD images were
bias-frame subtracted and flat-fielded corrected using twilight sky flats to
correct for pixel-to-pixel variations on the chip. No standard stars were
observed. Instead, we used secondary standards (comparison stars) in the field
near the target. The secondary standard stars were taken from \citet{reig15}.
Initially, we extracted the instrumental magnitudes of all the field stars
marked as standard stars in \citet{reig15}. However we checked for their
stability and intra-night small amplitude variations. From the entire list, we
chose the  most stable ones. 

The selection of the secondary standard stars was made as follows: for each
night and each star, we obtained the light curve, i.e. time versus instrumental
magnitude. Then we subtracted the average instrumental magnitudes from each
measurement. Thus we created an average-subtracted light curve that fluctuates
around zero. Then we computed the standard deviation of the residuals that
resulted from subtracting one comparison star from the average of the remaining
comparison stars. We eliminated the stars with the larger value of the standard
deviation. We repeated the procedure until we were left with three or four
comparison stars.  The overall standard deviation computed using the light
curves of all of the final comparison stars was in the range  2--7
millimagnitudes and was adopted as the accuracy of our photometry for the
corresponding observing run.
The final light curves that were used for the timing analysis were obtained by
subtracting  the  mean of the average-subtracted light curves of the selected
comparison stars and the average-subtracted light curve of the target.

\subsection{Frequency analysis}
\label{methods}

The frequency analysis has been performed with the rectified light curves. In the case of TESS photometry, most of the light curves present slow trends due to instrumental drifts, long term aperiodic or quasi-periodic variability of circumstellar origin and sudden brightenings and fadings induced by outbursts. All these signals convey a large number of frequencies that pollute the low frequency spectrum. To remove these variations we have rectified the light curves with fitting splines, by means of the Savitzky-Golay filter implemented in the {\tt Lightkurve} package. 

However, we have carefully inspected the original, non-rectified light curves in order to detect the occurrence of mid to long term quasi-periodic variability and light outbursts, and the relationship between these events and amplitude variations of the short-term variability, which could be related to mass-loss episodes. The results of this analysis are presented in Sects.~\ref{outbursts} and \ref{slow}.

We applied standard Fourier analysis and least squares methods to search for
periodicities in the light curves. In particular, we used the {\tt Period04} 
program \citep{lenz05} and PASPER \citep{diago08} as the basic analysis tools. 
These programs search for frequencies by means of standard Fourier analysis.
Once a frequency is detected, the program adjusts the parameters of a sinusoidal
function using a least-squares fitting and prewhitens the signal from this
frequency i.e. subtracts a synthetic sinusoidal light curve from the data in the
time domain. Then, a new frequency is found in a new step and the subsequent
least-squares fitting is performed, allowing the two frequencies to move in order
to get the minimum variance. The method is iterative and stops when the removal
of a new frequency is not statistically significant. We consider that a
frequency is statistically significant when the signal-to-noise (S/N) ratio is
above 4  \citep[see][for details]{breger93}. The signal is the amplitude of the
peak at the given frequency and the noise is taken as the average amplitude in
the residual periodogram after the prewhitening of all the frequencies detected
in a certain range around the peak frequency. We calculated the frequency
resolution following the Rayleigh criterion, i.e., $1/T$ , where $T$ is the time
duration of the light curve.

The choice of the range of frequencies around the peak frequency to compute the
noise affects the significance of the detection. Unfortunately, there is not a
unique good value as it should match the frequency range covered by the
periodogram but also because of the different types of frequency spectra. In the
presence of red noise (typically below 2 \cd), a wide interval may include low
amplitude noise above the region where the red noise dominates increasing
artificially the significance of the detection. A wide interval normally
excludes the low-amplitude high-frequencies peaks. On the other hand, a too
narrow range decreases the significance of the central frequency when it appears
in a group. Different authors have computed the average noise over different
size windows centered on the relevant frequency:  5 \cd, 
\citep{gutierrez-soto07a,semaan18}, 3 \cd \citep{papics11}, 2 \cd
\citep{labadie20b}, or even as low as 1 \cd
\citep{papics17,burssens19,szewczuk21}. For other ways of computing the
significance of a peak in a periodogram see \citet{reegen07}.

For the TESS light curves, we run our analysis twice, one over
the 0--20 \cd interval using an interval of 5 \cd around the peak frequency to
derive the noise and another one over 3--20 \cd with a 1 \cd  noise interval
width. The TESS light curves used in the analysis were created by sector. When the source was observed during more than one sector, we also analyzed the combined light curve of all contiguous sectors.

In the case of ground-based photometry, we analyzed the light curves created per observing period, which typically covers a few days (see Tables~\ref{logobs1}--\ref{logobs3}). In addition to using the light curves
made of the difference between the source and the average of the comparison
stars, we also performed a frequency analysis on the light curves made from a
combination of comparison stars, without the source. This way we can ensure the
reality of the periodicity found. Only if the periodicity detected on the source
disappeared in the light curves of the comparison stars, the periodicity was
considered to be a firm detection. The interval of frequencies to derive the
average noise was taken to be 5 \cd.

As an example of the frequency analysis, we show the results for 4U\,0115+63 in
Fig~\ref{4u0115}. We show the light curves, periodograms and phase diagrams for
the TESS and Skinakas observations. The most prominent feature of the
periodograms is a peak at frequency 3.3 \cd, that it is detected in both data
sets.  The results of the TESS and ground-based data analysis of the individual sources are reported and discussed in Appendix~\ref{indiv}.

\subsubsection{Independent frequencies}
\label{indfreq}

In densely frequency spectra, certain frequencies may result from the linear
combination of some basic frequencies \citep[see e.g.][]{kurtz15}. In this work,
we shall refer to these basic frequencies as independent or parent frequencies.
In order to identify the independent frequencies we applied the following
procedure \citep{papics12,papics17,szewczuk21}: First we check if two closely
spaced frequencies are separated by less than $2.5/T$, where $T$ is the total
duration of the light curve. This is known as the  \citet{loumos78} criterion,
which states that the minimum separation between two frequencies so that the
wings of one frequency do not affect significantly the other should be at least
2.5 the Rayleigh resolution ($1/T$). If that is the case, then we keep the peak
with the larger amplitude and discard the other peak. Second, we searched for
linear combination of the type

\begin{equation}
\label{lc}
f_i = m \times f_j + n \times f_k
\end{equation}

\noindent with integers $m$, $n$ from $-3$ to 3. If the combination on the right-hand side of eq.~(\ref{lc}) is less than $1/T$, then $f_i$ is considered to be a linear combination of $f_j$ and $f_k$. Some linear combination may
occur by chance. In order to asses whether a combination is real or not, we
computed two parameters. The first one is taken from \citet{szewczuk21}

\begin{equation}
P_1=\frac{\sqrt{A_j^2+A_k^2}}{A_H}
\end{equation}

\noindent where $A_j$ and $A_k$ are the amplitude of the parent frequencies and
$A_H$ is the highest amplitude of the spectrum. The smaller $P_1$ is, the
higher the probability that the resulting combination occurs by chance (hence the modulation is real). The second parameter is taken from \citet{papics12}

\begin{equation}
P_2=\frac{\sum{A_j}}{O_i}
\end{equation}

\noindent where $O_i=|n|+|m|$.  If $P_2>A_i$ then the corresponding peaks
are likely to come from a combination.

According to these two parameters, we shall consider as independent frequencies
those that do not satisfy eq~(\ref{lc}) as well as those that satisfy that
equation but have $P_1<1$ and $P_2<A_i$. By imposing these two conditions, we
minimize the risk of discarding frequencies whose linear combination occurs by
chance. To estimate the uncertainties in the frequency, we opted for the more
conservative approach of using $\sigma(f)=1/4T$
\citep{kallinger08}. 

%following the prescription in \citet[][see also
%\citealt{reegen07}]{kallinger08}, namely $\sigma_f=1/(T \sqrt{\rm{sig}})$, where
%${\rm sig}=\pi \log e/4 (S/N)^2$. }

\section{Results}

%\citep{walker05a,walker05b,gutierrez-soto08,huat09,neiner09,emilio10,balona11,szewczuk21}.

We have found that all BeXBs analyzed in this work show fast light intensity
variability. The amplitude of this variability from maximum to minimum within one night measured from the normalized light curves ranges from 0.02-0.06 mag in the
ground-based photometry and from 0.1\% to 5\% in flux
in the TESS band-pass.  In addition, all systems display multi-periodic variability,
although the amplitude of the modulations and number of frequencies in the
periodograms vary significantly for different sources. The highest number of
frequencies with $S/N>4$ was found for the SGXB Vela X--1 with 79 detected.
Among BeXB, IGR\,J01363+6610 was the source with the largest number of peaks
with 62.  However, as explained above, many peaks appear as linear combinations
of a small number of parent frequencies. A detailed analysis per source is given
in Appendix~\ref{indiv}. Appendix~\ref{tables-indfreq} (Tables~\ref{freq-tess1} to \ref{freq-sgxb}) gives the lists of
independent frequencies after the selection made according to
Sect.~\ref{indfreq}. 
while Appendix~\ref{ground-freq} gives the significant frequencies detected in the data obtained from the ground-based observatories.

%For the error in the frequency, amplitude, and phase values, we use the analytic
%formula from Montgomery and O\u2019Donogohue (1999).
%-------------------------FIG 2------------------------------------
    \begin{figure}
   \centering
   \includegraphics[width=10cm,clip]{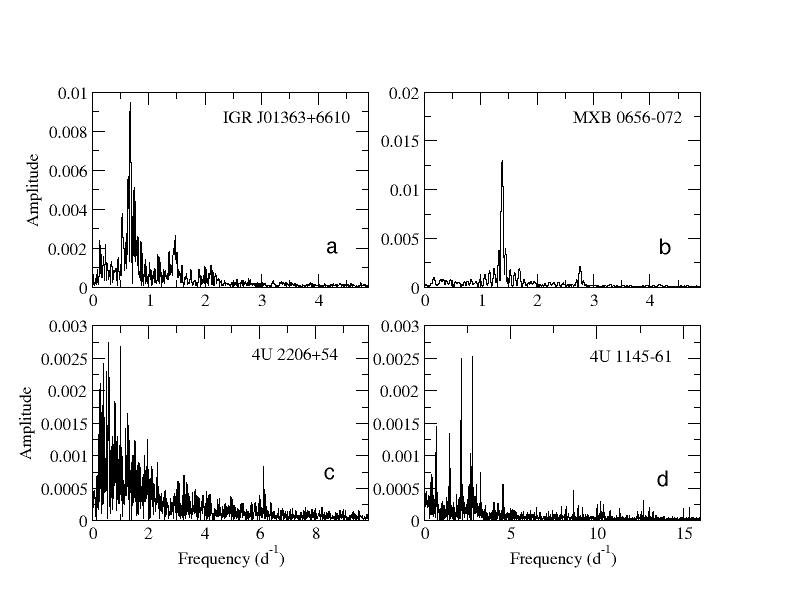}
      \caption{Representative examples of the different types of periodograms in BeXBs.}
	 \label{spec-type}
   \end{figure}

%_____________________________________________________________

\subsection{Periodograms}
\label{periodograms}

From a purely phenomenological point of view, the fast-time variability of BeXBs
can be classified into different categories depending on the characteristic
features of their periodograms \citep[see e.g.][]{labadie22}. We can identify four different types (Fig.~\ref{spec-type}):

\begin{itemize}

\item {\it Groups of frequencies} (Fig.~\ref{spec-type}$a$).  The most common pattern of variability in
the periodograms of classical Be stars is the existence of frequency groups 
\citep{walker05a,kurtz15,semaan18,baade18,labadie21,balona21}.  The most common configuration includes three groups at
frequencies $\simless0.5$ \cd, 0.5--3 \cd, and 1--6 \cd. Sometimes the first
group is absent. Regardless of whether two or three groups show up in the
periodogram, the central frequency of the highest-frequency group is almost
always about twice that of the second group. As in classical Be stars, frequency
groups are also present among BeXBs and represent the most common configuration
with 14 sources (67\%) of our sample of 21 BeXBs with TESS data. The
frequency spectra in this category show different degrees of complexity, but in
general about 70\% of them display  an approximate harmonic relationship between the second and third group. 

Typically, one of the two higher-frequency groups contains a
dominant sharp peak that stands above the broader and lower-amplitude peaks
that conforms the group. The sharp peak may appear in the second group, as in 1A\,1118--616 (Fig.~\ref{lc2}), in the third group, as in 1A\,0535+262 (Fig.~\ref{lc1}) and RX\,J0812.4--311 (Fig.~\ref{lc2}), or in both as in 4U\,1145--61 (Fig.~\ref{lc2}) and RX\,J0440.9+4431 (Fig.~\ref{lc1}).

\item {\it Isolated signals} (Fig.~\ref{spec-type}$b$). In about 18\% (4 sources) of the BeXBS with TESS data that we analyzed, the frequency spectrum shows one
or more isolated peaks. These sources are IGR\,J06074+2205 (Fig.~\ref{lc2}), MXB\,0656--072 (Fig.~\ref{lc2}), GX\,304--1 (Fig.~\ref{lc2}), and GRO\,J2058+40 (Fig.~\ref{lc3}),  In all cases, the most significant peak is below 2.5
\cd. 

\item {\it Stochastic variability} (Fig.~\ref{spec-type}$c$). Red noise is stochastic variability whose
amplitude decreases rapidly with increasing frequency. It may be formed by pure noise,
without any clear peak or by a forest of signals. Stochastic variability in early-type stars can be produced by inhomogeneities in the stellar surface or wind in combination with rotation \citep{aerts18,simon18} or by  interval gravity waves generated at the interface of the convective core and radiative envelope  \citep{bowman19b,bowman20}. This pattern is characteristic
of supergiant stars \citep{bowman19}  and can be clearly identified in the SGXBs 2S\,0114+65 (Fig.~\ref{lc3}), Vela X--1 (Fig.~\ref{lc3}), and
IGR\,J00370+6122 (Fig.~\ref{lc3}). Stochastic variability is sometimes a prominent feature in the periodograms of Be stars, where inhomogeneities in the inner disk may also contribute to the red noise. We have detected this kind of variability in 5 
%(24\%, compared to 34\% in \citet{labadie22} sample) 
of the BeXBs with TESS data. One interesting difference between the frequency spectra of SGXBs and the BeXBs in this group is that the BeXBS also show
isolated peaks at higher frequencies. This is notoriously evident in  4U\,2206+54 (Fig.~\ref{lc3}), RX\,J1037.5--5647 (Fig.~\ref{lc2}) and KS\,1947+300 (Fig.~\ref{lc3}). 

\item {\it High-frequency signals} (Fig.~\ref{spec-type}$d$). About half of the BeXBS in our TESS sample display narrow peaks at frequencies above 6 \cd. Typically, these peaks have lower amplitudes than those detected at lower frequencies but they are significant above the local noise level (i.e, S/N above 4). 
%This is somewhat surprising because, \citet{labadie22} reported that only 16\% of the early-type classical Be stars exhibited high-frequency signals. 

\end{itemize}

Figures~\ref{lc1} to \ref{lc3} show the TESS light
curves and frequency spectra of the sources analyzed in this work.

%------------------------FIG 3-------------------------------------
   \begin{figure}
   \centering
   \includegraphics[width=11cm,clip]{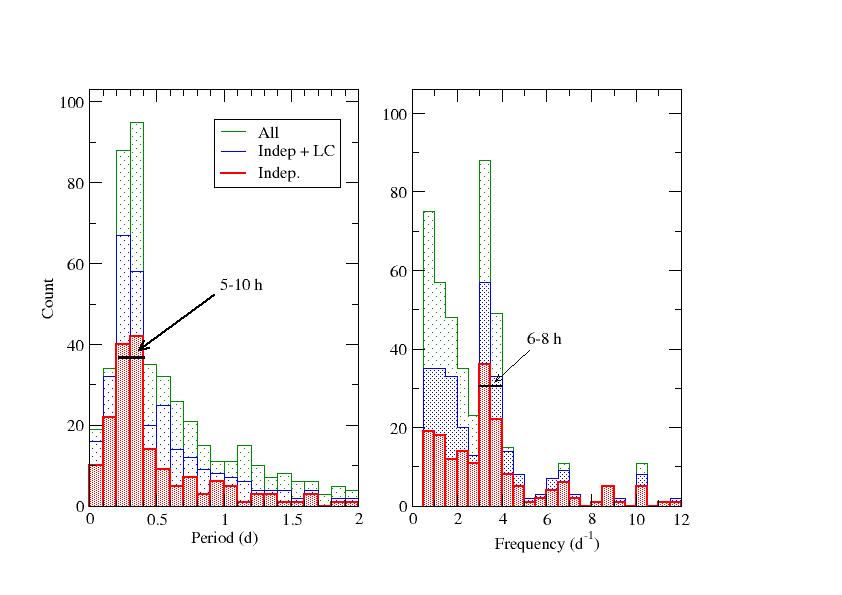}
      \caption{Distribution of periods (left) and frequencies (right) with {\it
      TESS}. The green line represents all detected signals, the blue line
      corresponds to remaining signal after removing closely spaced peaks, and
      the red line is final distribution of significant independent peaks.}
	 \label{per-freq_hist}
   \end{figure}

%-------------------------------------------------------------
%------------------------FIG 4-------------------------------------
   \begin{figure}
   \centering
   \includegraphics[width=10cm,clip]{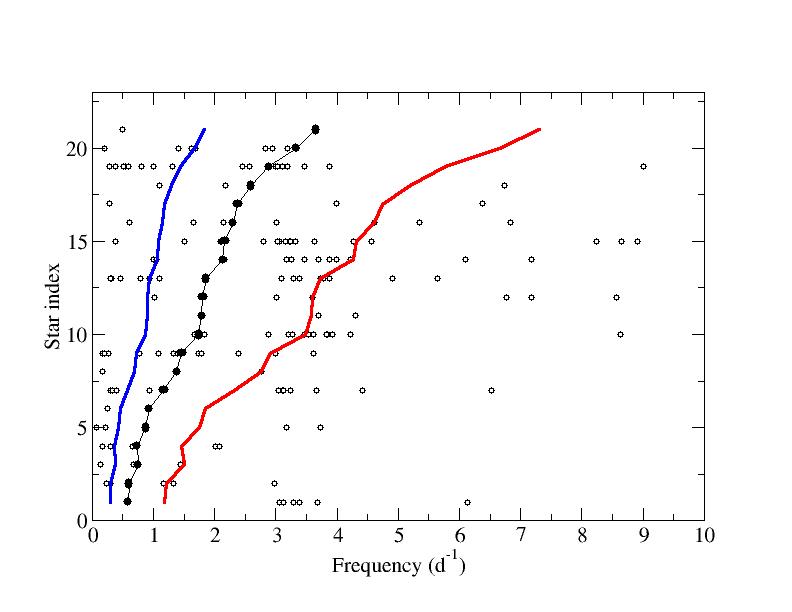}
      \caption{Independent frequencies detected in each source (empty circles), ordered by the strongest frequency (filled circles). The lines on both sides of the main trend correspond to 0.5 and 2 times the dominant frequency.}
	 \label{freq-ordered}
   \end{figure}

%-------------------------------------------------------------

\subsection{Distribution of frequencies}

Figure~\ref{per-freq_hist} shows the distribution of the periods (left) and
frequencies (right) of the signals detected in our sample of BeXBs. We use the word period in the general sense as the inverse of the frequency but it should be noted that it does not necessarily mean a stable signal with persistent variability. The green area
corresponds to all detected modulations, the blue histograms represent the
signals after the removal of closely spaced peaks, while the red lines represent
the parent (i.e. independent) frequencies. Because we are interested in the
variability from the Be star itself, we do not include periods longer than 2
days (or frequencies below 0.5 \cd), which most likely correspond to difference of higher frequencies \citep{baade18} or may be associated with stochastic variability or inhomogeneities in the disk. The distribution of periods is clearly dominated by
fast time variability. This type of variability in BeXBs occur mainly on time
scales faster than $\sim12$ hr. The peaks of the distributions correspond to
modulation of 5--10 hours. However, there is a significant number of periods
shorter than $\simless 4$ hr ($\sim6$ \cd). The frequency histogram shows a secondary maximum at lower frequencies, suggesting that variability on time scales $\sim 1$ day is also important in BeXBs.

The dominant frequencies of each star in the sample are shown in Fig.~\ref{freq-ordered}. In this figure, we ordered the systems taking into account the strongest frequency. We assigned a number sequentially from the lowest to the highest amplitude ($Y$ axis). The  filled black circles represent the strongest frequency, while the lines correspond to 0.5 and 2 times that frequency. We note that about half of the BeXBs display an  approximately 2:1 harmonic relation with respect to the dominant frequency (see circles over the red line in Fig.~\ref{freq-ordered}).

%------------------------FIG 5-------------------------------------
   \begin{figure}
   \centering
   \includegraphics[width=11cm,clip]{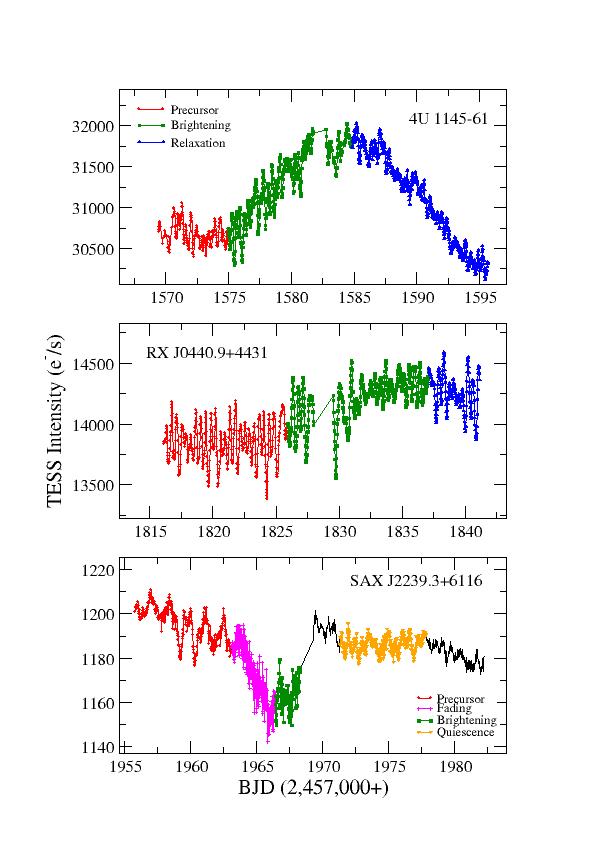}
      \caption{TESS light curves of three BeXBs that show light intensity outbursts: 4U 1145--61 (top), RX J0440.9+4431 (middle), and SAX J2239.3+6116 (bottom).}
	 \label{outlc}
   \end{figure}

%-------------------------------------------------------------

%------------------------FIG 6-------------------------------------
   \begin{figure}
   \centering
   \includegraphics[width=11cm,clip]{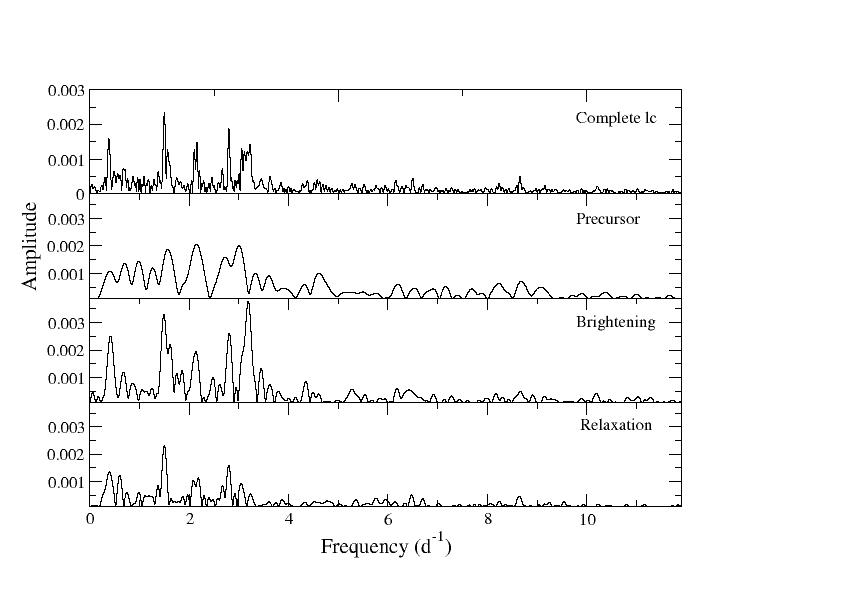}
      \caption{Frequency spectra of 4U\,1145--61 at different stages of the light outburst (see top panel in  Fig.~\ref{outlc}). }
	 \label{4u1145_dft}
   \end{figure}

%-------------------------------------------------------------
%------------------------FIG 7-------------------------------------
   \begin{figure}
   \centering
   \includegraphics[width=11cm,clip]{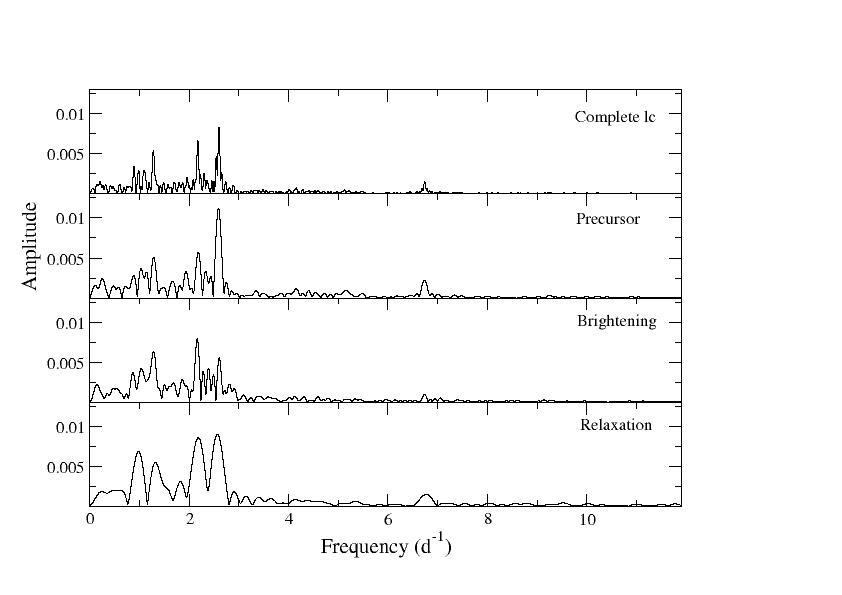}
      \caption{Frequency spectra of RX\,J0440.9+4431 at different stages of the light outburst  (see middle panel in  Fig.~\ref{outlc}). }
	 \label{ls4417_dft}
   \end{figure}

%-------------------------------------------------------------
%------------------------FIG 8-------------------------------------
   \begin{figure}
   \centering
   \includegraphics[width=11cm,clip]{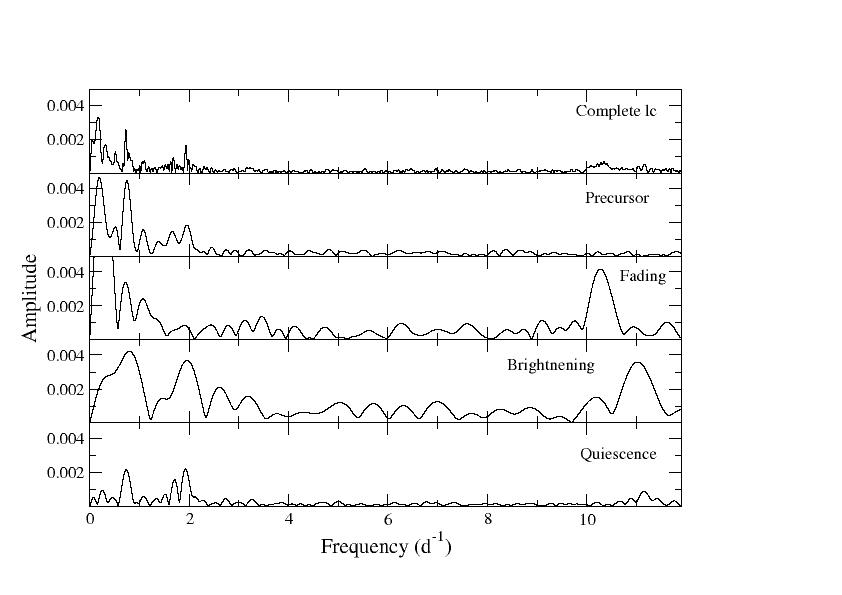}
      \caption{Frequency spectra of SAX\,J2239.3+6116 at different stages of the light outburst (see bottom panel in  Fig.~\ref{outlc}). }
	 \label{sax2239_dft}
   \end{figure}

%-------------------------------------------------------------

\subsection{Stars showing light outbursts}
\label{outbursts}

Although there is general agreement that the disk is fed from material from the photosphere of the Be star and not from an external source (as it would be in the case of an accretion disk), the details of how matter overcomes the gravitational potential of the massive star are not clear. 
Outbursts in Be stars are sudden increases or decreases of the light brightness by a few percent over a few days, followed by a return to the baseline brightness previous to the outburst. They are interpreted as episodes of mass transfer from the star to the disk, and are recorded as light as well as line emission variation \citep{rivinius98,huat09,rivinius16,semaan18,labadie22}.  For a detailed study of the effects of time variable mass loss rates on the structure of the disk and its observational consequences, the reader is referred to \citet{haubois12}.

Frequency analysis of Be stars have demonstrated that the variability observed at various phases of an outburst translates into different pulsational behavior. One of the observational consequences of these outbursts is the appearance and disappearance of certain frequencies at different stages of the outbursts and the transfer of amplitude between frequencies in the periodogram \citep{huat09}. Another strong observational difference is the larger oscillation amplitude during the rise and peak of the outburst compared to its decay \citep{semaan18}. Finally, the emergence of a group of frequencies during the brightening phase has been also attributed to outbursts \citep{huat09, labadie22}.

From the inspection of the no-rectified TESS light curves, outbursts as light brightening were detected in 4U\,0115+63, 1A\,1118--616, 4U\,1145--61 and RX\,J0440.9+4431. An outburst as a light fading was found in SAX\,J2239.3+6116. To study the temporal variations in the frequency spectrum we divided the light curves containing outbursts in four separated phases, as in \citet{huat09} and \citet{semaan18}, following the schematic picture proposed by \citet{rivinius98}, based on spectra. These phases are quiescence, precursor, brightening or fading, and relaxation. During the quiescent phase, the  mean flux is stable and variability is dominated by oscillations. The precursor phase corresponds to a slight decrease in flux prior to the rise or fade. The brightening or fading phase is when the flux gradually increases or decreases. During the relaxation phase, the star slowly recovers the phase of quiescence.

We have performed a frequency analysis for each phase of the outbursts separately, to study the possible variations of the pulsational behavior related to the outburst. However, the detailed analysis of the changes expected was hampered for the short time interval covered by TESS sectors, which translates into a low frequency resolution in the periodograms. 

In Fig.~\ref{outlc} (top panel) we present the TESS light curve during sector 10 in 4U\,1145--61, which can be separated into three phases (precursor, brightening, and relaxation) that would correspond to a light outburst. The low frequency resolution of the precursor spectrum does not allow a meaningful comparison with the other two phases. However, the frequency spectra during the brightening and relaxation phases were markedly different (Fig.~\ref{4u1145_dft}). Not only
the amplitude of the dominant frequencies at 0.4, 1.5, and 2.8
\cd exhibited larger amplitude during the brightening phase, but also a new frequency emerged at 3.2 \cd, which could be associated with the 2.8 \cd or be the harmonic of the 1.5 \cd oscillation.

The TESS light curve of RX\,J0440.9+4431 (sector 19) shows evidence for an outburst, where a precursor phase, a brightening or outburst phase, and a relaxation phase can be distinguished (middle panel in Fig.~\ref{outlc}). However, the short duration of the relaxation phase leads to a low-resolution frequency spectrum, which makes the comparison with other phases difficult (Fig~\ref{ls4417_dft}). We observed the appearance of a frequency at 1.9 \cd in the precursor phase that was not seen in the brightening or relaxation phases. The amplitude of the dominant peak in the complete light curve at 2.6 \cd decreased significantly during the brightening phase. It seems that part of the amplitude to that peak shifted to the second (in significance) peak, namely that at 2.2 \cd. During the precursor phase the ratio of the amplitudes was
$A_{2.6}/A_{2.2} > 1$. It reversed during the brightening phase, while it was $\sim$ 1 during the relaxation phase.

In the light curve (sector 24) of  SAX\,J2239.3+6116, the outburst appears as a light fading with a subsequent recovery of the mean brightness (bottom panel in Fig.~\ref{outlc}). The low resolution of the periodograms prevented a detailed analysis of the variations in the frequency spectra. However, a remarkable feature is the appearance of a strong high frequency signal at 10.5 \cd in the fading phase, which is completely absent in the precursor phase. In the relaxation phase this signal was still present with about half of the amplitude, but a new, strong signal appeared at 12 \cd (Fig~\ref{sax2239_dft}). Both high frequency peaks almost disappeared in the subsequent phase of quiescence.

For the remaining stars, the low resolution of the periodograms also prevented a detailed analysis. In 1A\,1118--616 the TESS light curve of sector 10 shows two short episodes of brightening and relaxation and a subsequent phase of quiescence. During the second brightening phase an oscillation at 2.1 \cd emerged, which was not present in the relaxation or quiescence phases. In 4U\,0115+63 even the presence of an outburst is uncertain, as it is very short and with a low amplitude, although a few low amplitude peaks are apparent in the brightening phase, at frequencies between 5 and 6 \cd, which were no longer present in the relaxation and quiescent phases. 

A common feature observed from space photometry is the appearance of frequencies about 10-20\% lower than a nearby strong stellar frequency during the episodes of mass ejection. They are interpreted as arising in the circumstellar environment due to an inhomogeneous distribution of the recently ejected material orbiting the star (\v{S}tefl frequencies, \citet{stefl98, baade16}). We have not detected this kind of frequencies in any of the BeXBs light curves presenting outbursts, although most probably the low resolution of our periodograms would have prevented their detection if they were present.  

On the other hand, \citet{labadie22} have found that all Be stars exhibiting outbursts also display one or more frequency groups, suggesting a strong link between these features. This is also the case with the five BeXB stars discussed in this section.

In summary, the fact that we observe similar features in the light curves of BeXBs and classical Be stars strengthens the view that they share the same mass-loss mechanism that lifts matter into an orbit where viscosity takes over to form a Keplerian disk. These common features are: the appearance and disappearance of frequencies, the shift in the amplitude between certain frequencies, the larger oscillation amplitude during the rise of the outburst and the emergence of groups of frequencies during the brightening phase.

\subsection{Stars showing slow variability}
\label{slow}

A common variability pattern of classical Be stars is the presence of mid to long term quasi-periodic variability, with time scales from a few days to months or even years. A relationship between the variations of the mean magnitude and the amplitude of the fast variability due to non radial pulsations has be seen in many Be stars: when the amplitude of the fast variability increases, the mean brightness of the star increases as well. Since the mean brightness is sensitive to the amount of matter in the inner disk, such correlations probably show that the in-phase superposition of NRP modes can provide a mechanism to lift matter into the circumstellar disk \citep{baade17,baade18,labadie21}. 

We have inspected all the original, non-rectified TESS light curves of our sources, in order to detect long-term variability and to study whether the relationship between slow and fast variability described in the previous paragraph holds for the donor stars in BeXB systems. Note that this kind of variability is different from the outbursting behaviour analysed in the previous subsection: we refer as long-term variability the smooth oscillations around the mean brightness, while outbursts are sudden increases or decreases of brightness followed by a return to the baseline flux.

Almost all stars in our sample present some kind of long-term variability. However, as the timescales of this variability are in general much larger than the time span of the TESS sectors ($\sim$27 days), in most cases we only observe a monotonous brightening or fading behaviour, which prevent us to compare different phases of the long-term variability with changes in the amplitude of the short-term variability. Only two stars present quasi-periodic variability with a timescale short enough to include several maxima and minima within a few consecutive TESS sectors. They are  1A\,0535+26 and X Per. In 4U\,1145--61 the shape of the light curve could be interpreted both as a long-term variability or as an outburst (see Fig.~\ref{outlc}). We have finally considered it as an outburst due to the marked differences between the frequency spectra at the different light curve phases, including the emergence of a new frequency, as described in the previous subsection. This behaviour is more characteristic of the outbursts than of the smooth long term variability.

\begin{figure}
   \centering
   \includegraphics[width=10cm,clip]{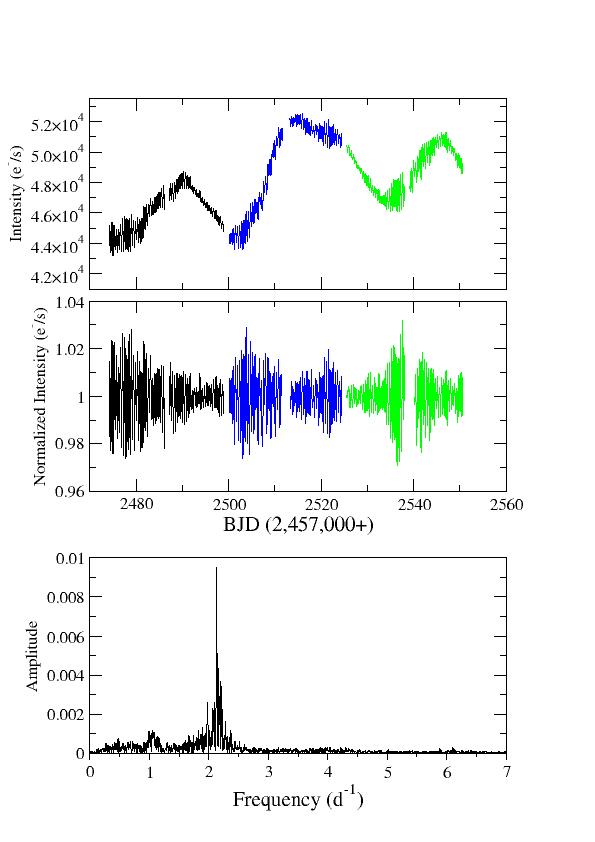}
      \caption{Top panel: Raw TESS light curve of 1A\,0535+26. We use different colors for the different consecutive TESS sectors in which the star was observed. Middle panel: Rectified light curve. Bottom panel: Periodogram.}
	 \label{1a0535_long}
   \end{figure}

1A\,0535+26 was observed by TESS in sectors 43 to 45. The light curve displays a quasi-sinusoidal long term variability, with and amplitude of about 15\% in flux (Fig.~\ref{1a0535_long}, top panel). The light curve also presents short term variability. The frequency analysis reveals the characteristic pattern of two frequency groups, centered around 1 and 2 \cd. The most significant frequency, at 2.13 \cd and with amplitude of about 1\%, belongs to the second group.

There is a clear correlation between the short and long term variability. The complete light curve includes three local maxima. It can be seen than the amplitude of the short term oscillations varies along all the light curve, with a fixed pattern with respect to the long term variability: the amplitudes of the former are higher after each local minima, and gradually diminish along the ascending branches up to the maxima. After each maxima the amplitude reach the lowest values.

Before the first maximum, the high amplitude of the oscillations corresponds to the in-phase superposition of frequencies in the first and, mainly, in the second group. Before the other two maxima the behaviour is slightly different. The highest amplitude oscillations at the beginning of the brightening phase correspond to in-phase superposition of frequencies of the second group. Later on, the amplitude of these frequencies decline, while frequencies of the first group present in-phase superposition. 

This behaviour is consistent with the general trend observed in classical Be stars where, as stated above, the mean brightness of the star starts to increase when the amplitude of the fast variability reaches its maximum value. 

\begin{figure}
   \centering
   \includegraphics[width=10cm,clip]{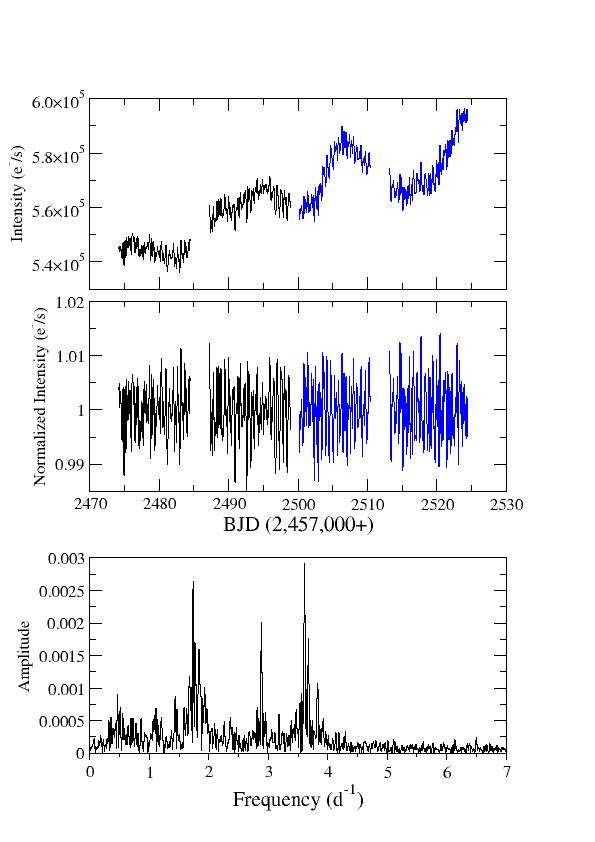}
      \caption{Same as Fig.~\ref{1a0535_long} for X Persei. }
	 \label{xper_long}
   \end{figure}

X Persei was observed by TESS in sectors 18, 43 and 44. In all of them it  presents long term variability. In Fig.~\ref{xper_long}, top panel, we display the light curve of sectors 43 and 44, which includes three local maxima and the rise to a fourth one. The frequency spectrum also presents the two frequency groups pattern, with groups at 1.8 and 3.6 \cd. In addition, there is a significant high frequency at 2.8 \cd, isolated from any group, which appears in sectors 43 and 44 but is not present in sector 18. However, unlike 1A\,0535+26, the frequency spectrum of X Persei does not present any significant variation along the different phases of the long-term light curve.

%------------------------FIG 9-------------------------------------

%-------------------------------------------------------------

%------------------------FIG 10-------------------------------------

%-------------------------------------------------------------

\section{Discussion}
\label{discussion}

We have performed the first systematic analysis of the short-term
variability of Be stars in BeXBs. The main goal of this project was to search for
periodicities in the range of a few hours to days in the optical counterparts to
BeXBs and see whether the patterns of variability agree with the expected
behavior of classical Be stars.  This is the third paper of a series to study
the global optical variability of BeXBs. The first paper of the series
investigated the properties of their long-term photometric variability
\citep{reig15} and the second paper of their spectral variability
\citep{reig16}. An important addition of this work with respect to the previous
ones is the fact that the use of space-based observations allowed us to expand
our sample of sources to the Southern hemisphere.

\subsection{Comparison with B and classical Be stars}

There are two main types of pulsating early-type stars: the slowly pulsating B
stars (SPB) and the $\beta$ Cephei variables. $\beta$ Cephei variables are a
group of early B multi-periodic pulsators which exhibits short-term variations 
of brightness, radial velocity and line profiles
\citep{sterken93,aerts03,stankov05}. They are characterized by temperatures in
the range $\log T_{\rm eff} \sim [4.25-4.50]$ (spectral types B0--B2.5),
luminosities $\log L_{\odot} \sim [4.25-4.50]$, and masses $\sim7-20 \msun$
\citep{burssens19}. Pulsation periods range from 0.067 to 0.35 days (3--15 \cd)
and are associated with low-order $p$-modes and/or $g$-modes.  SPB stars \citep{waelkens91,decat02} differ
from the $\beta$ Cephei in terms of temperature and mass, as they extend to the bottom of the B-type main sequence (B9,  $\sim2 \msun$). Recent models indicate that their instability domain extends up to the upper main sequence, and hence there is a large overlap with the $\beta$ Cephei instability domain \citep{miglio07}. The pulsation periods are typically 0.5--3 days (0.3--2 \cd) in SPB stars. These periods are interpreted as NRPs of high-order $g$-modes
\citep{decat02}. Most $\beta$ Cephei and SPB stars are multi-periodic, which
causes beating phenomenon with periods of weeks to months in $\beta$ Cephei and
months to years in SPBs. The identification of NRPs with the optical variability
of SPB and $\beta$ Cephei stars has long been recognized
\citep{osaki74,telting97b,schrijvers02,berdyugina03,telting06,pigulski08,degroote09,labadie20b}.
Due to the overlaping instability domains, several pulsating B-type stars have been identified to show both low-frequency and high-frequency pulsations. They are called SPB/$\beta$ Cep hybrid pulsators
\citep{chapellier06,decat07,handler09,balona11,balona15}.

Classical Be stars are expected to pulsate as they are placed in the HR diagram in the instability domains of the $\beta$ Cep and SPB variables. Indeed, Be stars have been found to show a higher level of pulsational variability than non emission-line stars of the same spectral types \citep{diago08}, and, more recently, the analysis of high precision photometric time series from space observatories have revealed that non-radial pulsations are ubiquitous among them \citep{baade20}. 

The most common feature of the frequency spectra of Be stars is the presence of frequency groups, as described in the previous section. Another common feature is the long-term quasi-periodic or irregular variability, which appears as low-frequency or stochastic signals in the frequency spectrum. Both these patterns are not present in the frequency spectra of the slowly rotating $\beta$ Cep and SPB stars. 

Isolated signals, both at low (lower than 3 \cd) and high (higher than 6 \cd) frequencies are also found in the periodogram of Be stars. The low frequency signals could be interpreted as high-order g-modes characteristic of SPB stars, while the high-frequency ones could be associated with $\beta$ Cep p-modes. However, the ascription of a particular signal to a g- or p-mode type pulsations based only in its frequency is very uncertain because  the observed frequency can differ significantly from the actual frequency in the co-rotating frame of the star when the rapid rotation of Be stars is taken into account \citep{cox84}.

%The observed frequencies $f_{obs}$ are related to the frequencies in the co-rotating frame $f_{corot}$ by the expression:
%\begin{equation}
%f_{obs}=|f_{corot} - m \Omega |
%\end{equation}
%\noindent where $\Omega$ is the rotational frequency of the star and $m$ the azimuthal order of the pulsation mode. 

Stochastic variability is also present in the frequency spectra, as has been pointed out for some Be stars observed from space \citep{baade16}. Stochastic signals of astrophysical origin are induced in the frequency spectrum by non periodic variability, and it is strongest at the lower frequencies.

Another common feature of the Be stars' variability are the outbursts, sudden increases or decreases of brightness which are interpreted as associated with episodes of mass transfer from the star to the circumstellar disk. 

The present work reveals that BeXBs also display all this kind of variability. To compare the nature of the BeXBs variability with that of classical Be stars we will confront our results of the previous sections with those obtained by 
\citet{labadie22} for a larger sample of Be stars observed by TESS during the first year of the mission. In this comparison, we take into account that the spectral type distribution in BeXB lies in the narrow range O9-B2, with a peak around B0 \citep{negueruela98b,reig17}, and hence we compare our results with those of the early group (B3 and earlier) defined by \citet{labadie22}.

As stated above, one of the main characteristics of the Be stars' variability is the presence of frequency groups. Although several interpretations have been put forward to explain the existence of these groups, the favoured explanation is that the second group is normally associated with NRP, while the first and third groups are built from differences and sums and/or harmonics \citep{baade18b}. Alternatively, the groups might result from non-coherent variations associated with photospheric inhomogeneities or gas clouds. In this case, the mean frequency of the intermediate group would correspond to the rotation frequency, and the third group to its first harmonic \citep{balona20,balona21}.
In our sample of BeXBs we found this pattern to be present in 14 out of 21 sources studied, which means 67\% of the sample. 
In the sample studied by
\citet{labadie22}, about 91\% (199 of 218) of the early-type stars (earlier than B3)
displayed two or three frequency groups. Among the systems with frequency groups, we found that about 70\% of those BeXBs showed the typical configuration of two or three groups with the third group approximately located at twice the frequency of the second group. This compares to 85\% in \citet{labadie22}.

%The common configuration of three groups, with the central frequency of the highest-frequency group being about twice that of the second group, characteristic of classical Be stars (85\% in the sample studied by \citet{labadie22}), is also found in our sample ($\sim$70\%).

Four stars in our sample (18\%) contain isolated signals which do not belong to frequency groups. In five stars (24\%) we found stochastic variability. These figures compare with the 22\% and 33\% respectively found by \citet{labadie22} in their early group for the same kind of variability.

About half of the BeXBS in our sample display significant, isolated peaks at frequencies above 6 \cd. In this case our results differ from those of \citet{labadie22}, who only found this pattern present in the 18\% of their early type sample. This is the only feature in which our sample differs from the main values for classical Be stars as obtained by these authors. 

We detect light outbursts in five stars in our sample (24\%), to be compared with the 30\%  found by \citet{labadie22}. These authors also found that all system exhibiting outbursts (called flickers) present frequency groups, suggesting a strong link between these features. In our case, all five stars displaying light outbursts also present the frequency groups pattern.

Almost all BeXB stars in our sample display long-term variability. In two of them, 1A\,0535+26 and X Per, several maxima and minima are included in a few consecutive TESS sectors, which allowed us to compare the phases of the slow variability wit the amplitudes of the fast oscillations. In 1A\,0535+26 there is a clear correlation between the short and long term variability: the amplitude of the fast oscillations reach their maximum value immediately after each local minima, when the mean magnitude starts to increase, and diminish up to their minimum value after the maxima, when the star brightness fades. This behaviour has been observed in several classical Be stars (see, for instance, Appendix B in \citet{labadie21}), and can be interpreted in terms of that non radial pulsations at its maximum amplitude somehow lifts matter into the disk, increasing the mean brightness of the star. For X Per, however, there is not any detectable variation of the fast oscillations amplitude along the different phases of the long-term variability.  

Hence we can conclude that the variability patterns of the BeXBs are the same that those of classical Be stars, and both populations are indistinguishable in terms of pulsational characteristics. As the Be stars in BeXBs systems are the product of mass transfer in a close binary, our results would indicate that the structure of a post-mass transfer star is similar to that of an isolated star of similar mass. Or, alternatively, that most -or all- classical Be stars underwent mass transfer episodes in their past.

\subsection{BeXBs and SGXBs}

%--------------------------FIG 11-----------------------------------
   \begin{figure}
   \centering
   \includegraphics[width=11cm,clip]{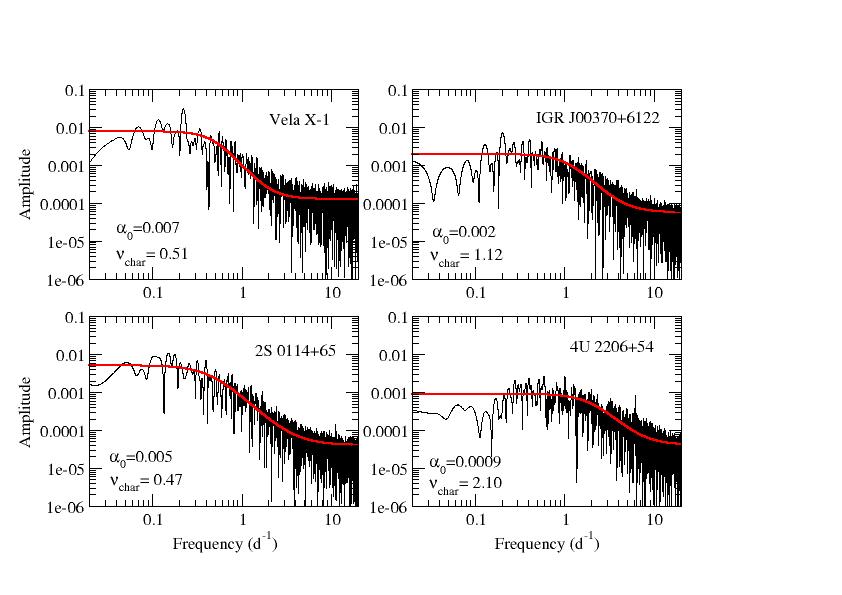}
      \caption{Comparison of the frequency spectra of the SGXBs with
      4U\,2206+54, a wind-fed BeXB. The red line is the best fit with the function given in eq.~(\ref{func}). }
	\label{BeXB-SGXB}
   \end{figure}
%_____________________________________________________________
%-----------------------------------------------------------------------------------------
\begin{table*}
\caption{Stable signals. This table gives the frequencies and periods of signals
that are detected at two or more different epochs. Errors in frequency and period are typically $\simless0.01$ \cd and $\simless 0.05$ hr, respectively (see Tables~\ref{freq-tess1} to \ref{freq-tess5}).}
\label{stable}
\center
\begin{tabular}{llll}
\hline
\hline
System			&Frequency		&Period		&Epochs		\\
			&(\cd)			&(hour)		&	\\
\hline		
4U\,0115+63 		&3.33			&7.2		&2008,2009,2018,2020  \\
4U\,0115+63 		&1.7			&14.1		&2009,2018  \\
RX\,J0146.9+6121	&2.9, 9.7		&8.3, 2.5	&2005-07,2018  \\
X\,Per			&3.61, 3.83		&6.6, 6.3	&2019,2021	\\
RX\,J0812.4--3114	&2.88			&8.3		&2019,2021	\\
GRO\,1008--57		&2.98, 10.19		&8.05, 2.35	&2019,2021	\\
RX\,J1037.5--5647	&0.31, 4.91, 5.64	&77.4, 4.89, 4.26&2019,2021	\\
1A\,1118--616		&0.93, 3.24, 4.42, 6.53	&25.8, 7.4, 5.4, 3.7&2019,2021	\\
4U\,1145--61		&2.16, 4.57, 8.65	&11.1, 5.25, 2.77&2019,2021	\\
GRO\,J2058+42       &2.37           &10.13          &2005,2019 \\
SAX\,J2239.3+6116   &0.73           &32.9              &2019,2020 \\
SAX\,J2239.3+6116   &1.93           &12.4             &2011,2020 \\
\hline
\end{tabular}
\end{table*}
%-----------------------------------------------------------------------------------------

The main physical difference between SGXBs and BeXBs is the way in which mass
is transferred between the optical donor and the neutron star. Supergiants
have strong stellar winds, while in BeXBs the main source of matter available
for accretion is the circumstellar disk. A visual inspection of the frequency spectra of BeXB and SGXB (see Fig.~\ref{BeXB-SGXB}) reveals that the different luminosity class of the mass donor star leads to a different characteristic pattern of the frequency spectra between BeXBs and SGXBs. The periodograms of the three SGXBs analyzed in this work are characterized by stochastic low-frequency variability and the lack of significant oscillations above $\sim2$ \cd. The stochastic low-frequency variability shows up as red noise, that is, as noise whose amplitude decreases towards higher frequencies. On top of the red noise component, coherent oscillations may also be present in the lower luminosity supergiant stars \citep{bowman19}. The stochastic low-frequency variability in supergiant stars has been explained as the interaction between internal gravity waves triggered by core and/or envelope convection and the stellar wind \citep{aerts17,bowman19}.

The peculiar system 4U\,2206+54 represents an interesting case for the comparison of the pulsational properties between BeXBs and SGXRs.
4U\,2206+54 is the only (together with LS\,5039, for which there are no TESS data available) permanent wind-fed HMXB with a main-sequence donor
\citep{ribo06}. The frequency spectrum of 4U\,2206+54 also shows a red noise
pattern. 
%Therefore, we conclude that a strong red noise component in the frequency spectra implies the presence of a stellar wind. This is
%an important result because it allows us to distinguish among the different
%types of X-ray accreting pulsars from optical variability studies. 
There are however two interesting differences between wind-fed BeXBs and the most
luminous supergiants 2S\,0114+64 and Vela X--1. The first difference is the fact
that the frequency spectrum of the supergiant systems, including the less
luminous system IGR\,J00370+6122 is completely featureless above 2 \cd,
whereas 4U\,2206+54 shows a significant isolated peak at 6.14 \cd, which is reminiscent of
a $p$-mode NRP.  This difference is best illustrated in Fig.~\ref{lc3}.
The lack of isolated high-frequency peaks in OB-type supergiant systems is expected, as the instability domains in the HR diagram for p- and g-mode NRP are restricted to the main sequence \citep{miglio07}.

%is an interesting result. Since the photospheric variability propagates into the wind, perhaps the stronger wind in supergiant stars dampens out the oscillations from the photosphere. This mitigated effect would suppress the high-frequency modulations because the intrinsic variability of the  wind is expected to have similar frequencies. In contrast, low-frequency oscillations involve longer time scales that would survive its journey across the wind. The weaker stellar wind in 4U\,2206+54 is partly transparent to the photospheric variability and we detect $\beta$ Cephei pulsations.

Another example is RX\,J1037.5--5647. This BeXB has never shown a large X-ray outburst. It has been observed at different X-ray luminosity levels ranging between $10^{34}$ and $3 \times 10^{35}$ erg s$^{-1}$ \citep{reig99,cusumano13}, which is typical of wind-fed accretors. The spectral type of the primary is B0 III-V \citep{motch97}. 
%hence we do not expect the stellar wind to completely veil the pulsations coming from the photosphere. 
Indeed, this source shows a mixed pattern in its periodogram with red noise and isolated high-frequency peaks. 

The lack of isolated high-frequency peaks in supergiant systems is statistically a sound result. \citet{bowman19} assembled a sample of 167 hot massive stars (91 supergiants and 76 stars with luminosity class III-V or unknown luminosity) with the purpose of investigating the incidence of coherent modes and stochastic variability.
Only two supergiant systems showed $\beta$ Cephei like pulsations, while all (except one) type Ia supergiants (29 systems) exhibited a frequency spectrum consistent with internal gravity waves (i.e. red noise) without any other kind of modulation. Only one system presented a modulation that could be attributed to rotation by a clumpy aspherical wind.
%Whether the explanation for this phenomenology lies in the interior of the stars (i.e. it is structural) or due to the interplay with the stellar wind remains an open question.

The second difference between BeXBs and SGXBs is illustrated in
Fig.~\ref{BeXB-SGXB}. This figure shows the frequency spectra of the three SGXB
and 4U\,2206+54 in log-log scale. We note that the red noise extends to higher
frequencies in the wind-fed BeXB 4U\,2206+54. To assess this effect we fitted the spectrum with the function \citep[see][]{bowman19}

\begin{equation}
\label{func}
    \alpha(\nu)=\frac{\alpha_0}{1+(\frac{\nu}{\nu_{\rm char}})^\gamma}+C_{\rm w}
\end{equation}

\noindent where $\alpha_0$ is the amplitude at zero frequency, $\gamma$ is the logarithmic amplitude gradient, $\nuÇ_{\rm char}=1/(2\pi\tau)$ represents the characteristic frequency, which is the inverse of the characteristic timescale, $\tau$, of the
stochastic variability, and $C_{\rm w}$ is a white noise term. The solid red line in Fig.~\ref{BeXB-SGXB} represents the best fit. The
characteristic frequency $\nu_{\rm char}$ is given in the bottom left corner of each plot. The luminous supergiant systems Vela X--1 and 2S\,0114+65 show larger amplitude at low frequency (higher parameter $\alpha_0$) but this amplitude falls down faster (lower parameter $\nu_{\rm char}$) than in the wind-fed main-sequence system  4U\,2206+54. The less luminous SGXB IGR\,J00370+6122 adopts intermediate values. The amplitude at zero frequency and characteristic frequency for RX\,J1037.5--5647 are $\alpha_0=0.0009$ and $\nu_{\rm char}=2.01$, in good agreement with 4U\,2206+54. This result agrees with the findings of \citet{bowman19}, who found a correlation between the luminosity of the star and the characteristic frequency of their amplitude spectra: the more luminous the source, the lower the characteristic frequency.

We note that the usage of equation~(\ref{func}) assumes that there is no low-frequency contribution from the disk. However, in Be stars, the inner disk often makes a contribution to the red noise. The fact that even with this contribution the amplitude of the low-frequency noise is higher in supergiant systems strengthens the differences between BeXBs and SGXBs.

\subsection{Comparison with previous results and stability of the pulsations}

In this section we compare our results from the ground-based photometry with
those of the TESS analysis and also with previous work. Also, because some
sources were observed by TESS at two different epochs, we can assess the
stability of the detected modulations over some period of time.  
TESS observed the southern ecliptic hemisphere during Year 1  (July 2018--July
2019) and during Year 3 (July 2020-July 2021) of the mission and the northern
hemisphere during Year 2 (July 2019-July 2020) and partly during Year 4 (July
2021-September 2022) of the mission.   Table~\ref{stable} gives the frequencies
and periods of signals that are detected at two or more different epochs
separated in time by at least two years. All systems observed in more
than one epoch show at least one stable signal.

We found only three reports on short-term photometric periodicities of BeXB in
the bibliography, corresponding to  LS~I~+61235/RX\,J0146.9+6121 \citep{sarty09}, GRO\,J2058+42 \citep{kiziloglu07b},
and 4U\,2206+54 \citep[][see also \citealt{hintz09}]{bugno09} based on  ground-based observations.

Observations performed during 2005 to 2007 revealed  three
frequencies at 2.914, 1.523 and 9.679 \cd in the $V$ band light curve of LS I
+61235/RX\,J0146.9+6121  with an estimated uncertainty of $1/4T\sim 0.008$ \cd \citep{sarty09}. The 2.91 and 9.67 \cd modulations were
interpreted as NRPs, while the 1.52 \cd (0.67 d) period as the rotation of the Be
star. We detect the same two frequencies attributed to NRPs in our observations
obtained 11 years later (2018). Because the time span of our ground-based observations was
relatively short (4 days), we could not firmly detect the 1,5 \cd frequency. Nevertheless, we detect a modulation at 1.4 \cd, which might be identified with the 1.5 \cd peak.  

\citet{kiziloglu07b} performed a time series analysis of the light curve of
GRO\,J2058+42 obtained from the ROTSEIIId robotic reflecting ground-based telescopes in 2005 and 2006 over 476 days. The light curve
contained a total of about 1440 data points. They found two closely spaced significant frequencies at 2.371 and 2.404 \cd, with an estimated error of $1/4T\simless 0.001$ \cd. These modulations were attributed to NRPs. Our
ground-based observations, which were taken in July 2020, also revealed a
modulation at $2.406\pm0.007$ \cd. Interestingly, a $2.373\pm0.011$ \cd frequency was also detected by TESS in August 2019, although we point out that the data would not permit us to resolve the 2.371 and 2.404 c/d as separate frequencies.

We note that the circumstellar disk in both RX\,J0146.9+6121 and GRO\,J2058+42
during the older observations were smaller than during our observations, as
indicated by the strength of the \ha\ line: $EW(H\alpha)\approx -(4-5)$ \AA\ and
$EW(H\alpha)\approx -(0-2)$ \AA\ for RX\,J0146.9+6121 and GRO\,J2058+42 during the
2005-2207 and 2010 observations, respectively. These values compare to
$EW(H\alpha)\approx -11$ \AA\ and $EW(H\alpha)\approx -10$ \AA\ during our
observations (P. Reig, unpublished). In fact, the 2005 observations of
GRO\,J2058+42 coincided with almost the complete dissipation of the disk.
Therefore, we conclude that the oscillations studied here are not related
to the disk and favour a photospheric origin.

4U\,2206+54 was observed for 22 nights during the Summer 2008 (HJD
2454631--2454680) by \citet{bugno09} and detected a modulation at 
$2.5726\pm0.0005$ \cd.
We confirm the presence of this periodicity in our ground-based observations in 2008. In this case,
our observations are contemporaneous to those reported by these authors. Thus we
cannot assess the stability of the pulsation over a long period of time.
However, the fact that two different data sets obtained from different sites and
instrumentation find the same frequency adds significance to the detection.

4U\,0115+63 shows a very strong modulation at 3.33 \cd that it is detected in
data spanning more than 12 years. This frequency is detected not only during the
two TESS epochs, but also during the ground-based observations taken in
2009 (see Fig.~\ref{4u0115}).

\section{Summary and conclusions}

This study has revealed that short-term variability is a very common if not 
ubiquitous feature intrinsic to BeXBs.  Although the detection of
modulated signals of the order of hours in early-type stars is not a rare 
phenomenon, this is the first time that a systematic study of the optical
companions in BeXBs is performed. Our findings can be summarized as follow:

\begin{itemize}

	\item We have detected short-term variability of the order of minutes to
	hours in the light curves of the optical counterparts to all the 
	high-mass X-ray binaries that we have analyzed.
	
	\item The fact that all sources display multi-frequency oscillations and that some of the
	detected frequencies of the modulations are higher than the
	maximum allowed for rotation favors the interpretation that
	non-radial pulsations is the main driver of the fast time optical variability in BeXBs. 
	
	\item The distribution of frequencies in BeXBs shows a peak at around 3--4 \cd (Fig.~\ref{per-freq_hist}), but there is also a significant contribution at lower frequencies.

    \item We have found that BeXBs and classical Be stars are indistinguishable in terms of pulsational characteristics. As in classical Be stars, BeXBs display frequency groups, long-term irregular variability, isolates signals, and light outbursts. 
    
    \item The light intensity variability of the Be star in BeXBs is also similar to that of classical Be stars. We have found both light outbursts and slow light variations (weeks to months). Light outbursts are characterized by sudden increases or decreases of the light brightness by a few percent over a few days, followed by a return to the baseline brightness previous to the outburst. These changes involve the transfer of power between certain frequencies as well as the appearance and/or disappearance of frequencies. Slow variability consists of smooth oscillations around the mean brightness.
    
	\item In contrast, SGXBs and BeXBs can be distinguished by the shape of their
	frequency spectra. The periodograms of SGXBs are characterized by significant red noise and the lack of pulsations at high frequencies. 
	
	\item When data from different epochs are available, some modulations show stability on time scales of years. These stable signals cover a wide range in frequency. While a few could be consistent with the rotation of the Be star, the majority are too low or too high to have a rotational origin and demonstrate that pulsations can also be long lived.

\end{itemize}

\begin{acknowledgements} 

We are grateful to the referee Dr. D. Baade for his useful comments and suggestions which improved the original version of this work. This research has made use of the SIMBAD database, operated at CDS, Strasbourg, France and of NASA’s Astrophysics Data System operated by the Smithsonian Astrophysical Observatory. This research made use of Lightkurve, a Python package for Kepler and TESS data analysis (Lightkurve Collaboration, 2018). Funding for the TESS mission is provided by NASA's Science Mission directorate.
IRAF is distributed by the National Optical Astronomy
Observatories, which is operated by the Association of Universities for
Research in Astronomy, Inc. (AURA) under cooperative agreement with the National
Science Foundation.
We thank observers K. Triantafyllaki, H. Psarakis, G. Savathrakis and A.
Tzoubanou and technical staff A. Kogentakis and V. Pantoulas for helping with
the Skinakas observations. Skinakas Observatory is a collaborative project of
the University of Crete and the Foundation for Research and Technology-Hellas.
The Aras de los Olmos Observatory (OAO) is a facility of the Astronomical
Observatory of the Valencia University (Spain). We thank O. Brevi\'a and V.
Peris for their support to the OAO observations.  The work of J.F. has been funded 
by the project PID2019-109592GB-100/AEI/10.13039/501100011033 from the Spanish
Ministerio de Ciencia e Innovaci\'on - Agencia Estatal de Investigaci\'on,
and by the Generalitat Valenciana project of excellence Prometeo/2020/085.

\end{acknowledgements}

\bibliographystyle{aa}
\bibliography{../../artBex_bib}
%\begin{thebibliography}{../artBex_bib}

%\end{thebibliography}

\begin{appendix}

\section{Results of the frequency analysis on individual sources}
\label{indiv}

In this section we present the results of our frequency analysis performed on each individual system. The sources are listed and ordered by right ascension.\\

\begin{figure*}
   \centering
   \includegraphics[width=17cm,clip]{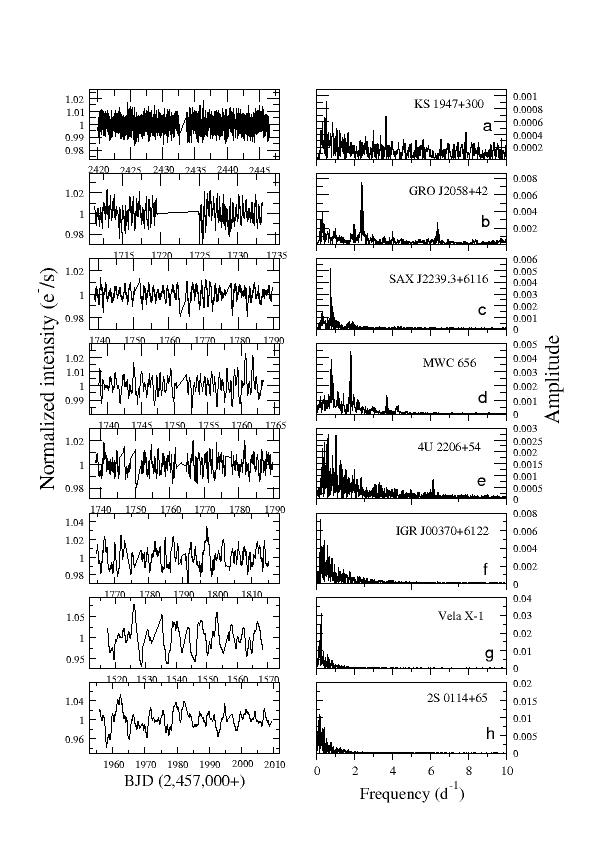}
      \caption{Light curves and periodograms, including the SGXBs Vela X--1, 2S\,0114+65, and IGR\,J00370+6122. The time in the light curves is the Barycentric Julian Date with reference time 2,457,000.}
	 \label{lc3}
   \end{figure*}

{\it IGR J00370+6122} (Fig.~\ref{lc3})\\

IGR\,J00370+6122 is a SGXB that was discovered in December 2003 during a 1.2 Ms INTEGRAL
observation \citep{denhartog06}. Based on high-resolution spectra taken at
different epochs, \citet{gonzalez-galan14} classified the optical counterpart BD
+60$^{\circ}$ 73 as a  BN0.7 Ib low-luminosity supergiant located at a distance
$\sim3.1$ kpc, in the Cas OB4 association.  \citet{denhartog06} and
\citet{intzand07} reported periodically variable orbitally modulated hard X-ray emission that
occurred with a period of 15.66 d and identified a possible 346 s period with the
spin of a pulsar companion. Orbital solutions are given by
\citet{gonzalez-galan14} and \citet{grunhut14} who performed radial velocity
measurements of a large number of optical spectra.

IGR\,J00370+6122 was observed by TESS during sectors 17 (October 2019), 18
(November 2019), and 24 (April 2020). The periodogram  is featureless above 2
\cd. Below this frequency, the frequency spectrum is characterized by a forest of
peaks (up to 38 frequencies above $S/N=4$ are detected in sectors 17 and 18) and
strong red-noise. In sector 24, an isolated low-amplitude (0.07\% in flux) peak
at 3.1 \cd is also detected. The stochastic nature of the variability where
multiple signals exist in the vicinity of each other is typical of supergiant
X-ray binaries. \\

{\it 2S\,0114+65} (Fig.~\ref{lc3})\\

This is one of the three supergiant X-ray binaries (SGXB) that we included in our
target list in order to compare the pattern of variability of BeXBs and SGXBs (the other two being IGR J00370+6122 and Vela X--1).
2S\,0114+650 was discovered in 1977 by the {\it SAS-3} Galactic
survey \citep{dower77}. The compact object is one of the slowest
rotating pulsars with a spin period of 2.7 hours. The donor star is LS~I+65$^{\circ}$010, a luminous B1 Ia supergiant \citep{reig96}. The orbital period
determined from  radial velocity measurements  is 11.6 days with an orbital
eccentricity $e = 0.18$ \citep{crampton85,koenigsberger06}.

2S\,0114+65 was observed by  TESS in sector 18 (November 2019), 24 (April
2020), and 25 (May 2020).  The periodogram displays strong red noise at low
frequencies. Below 2 \cd, there is a forest of signals, while above that
frequency no peaks are detected. This type of stochastic variations is a defining
characteristic of supergiant stars \citep{bowman19}. \\

\begin{figure*}
   \centering
   \includegraphics[width=17cm,clip]{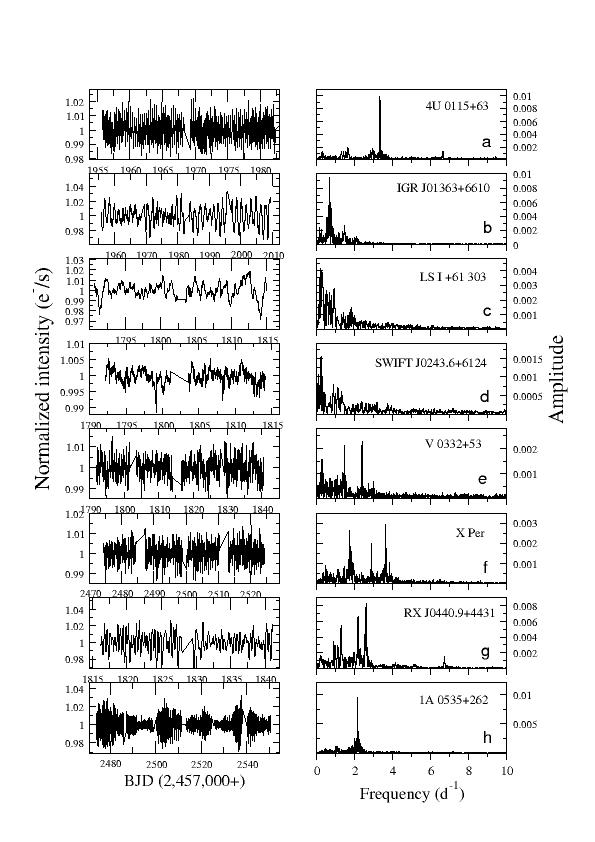}
      \caption{Light curves and periodograms. The time in the light curves is the Barycentric Julian Date with reference time 2,457,000.}
	 \label{lc1}
   \end{figure*}

{\it 4U\,0115+63} (Fig.~\ref{lc1})\\

4U\,0115+63 is one of the best studied BeXB, with  X-ray outbursts and disk
formation and dissipation phases every 3--5 years \citep{reig07b}. It harbors a
fast spinning neutron star, $P_{\rm spin}=3.6$ s, in a relatively narrow orbit,
$P_{\rm orb}=24.3$ days \citep{raichur10}. 

 \vcas\ was observed from the Skinakas observatory in four different epochs: two in
2008 and two in 2009.  Large amplitude variations ($\Delta V \simless 0.06$ mag)
were observed in virtually all the individual light curves obtained in a single
night (Fig.~\ref{4u0115}).  Although the nights of 15-16 September 2008 were not
photometric, a modulation of frequency $\sim$3.50 c d$^{-1}$ with an amplitude
of $\sim$ 15 mmag was tentatively detected. The star was observed again on 15 and 16
October 2008, and a frequency at 3.39 \cd\ was detected in the periodogram with
an amplitude of 18 mmag. 

In 2009, \vcas\ was observed from Skinakas on 22-27 September and 4-11 October.  A
periodogram obtained with these light curves clearly shows a dominant peak at
frequency 3.34 \cd with an amplitude of  17 mmag in September and 3.35 \cd with
an amplitude of  12 mmag in October. The analysis of all the 2009 data together
(the most complete set of observations)  confirms the periodicity at 3.33 \cd.
After prewhitening with this frequency, a peak at 1.73 \cd remains in the
periodogram with a $S/N$ slightly above the threshold of 4.

The 3.3 \cd modulation is clearly detected by TESS in observations made
during sectors 18 (November 2018), 24 (April 2020), and 25 (May 2020). The
amplitude of this peak ($\sim1$\% in flux) is one of the largest detected among all the BeXBs
analyzed in this work. An harmonic of this frequency at 6.60 \cd and a
subharmonic at 1.63 \cd are also detected in all three sectors. The structure
around the main peak at 3.30 \cd is similar to the window function, and hence
there is no evidence of frequency groups, at least in sectors 18 and 24.
In contrast, in sector 25 two groups of frequencies emerged, centred around 1.5 and
3.0 \cd, in addition to the 3.30 \cd peak and its harmonics.  
%If we interpret the single peak as a pulsation mode and the groups as rotational modulation, the
%rise of the frequency groups would indicate the appearance of inhomogeneities in
%or near the photosphere of the star. However, there is not any apparent change
%of the shape of the light curve which would indicate significant changes in the
%stellar photosphere. 
During sector 18, a peak at 1.7 \cd remains after filtering out for linear combination of frequencies. This modulation agrees with the one detected in the 2009 ground-based observations.\\

{\it IGR\,J01363+6610} (Fig.~\ref{lc1})\\

IGR\,J01363+6610 was discovered by the {\it INTEGRAL} imager IBIS/ISGRI on April 19, 2004
during observations dedicated to the Galactic Plane Scan \citep{grebenev04}. Optical observations
by \citet{reig05a} found an H$\alpha$ emission star located within the INTEGRAL uncertainty circle and determined the spectral type to be
B1IV-Ve. Although the nature of the compact object has not been resolved, the
X-ray analysis performed by \citet{tomsick11} confirmed that the system is a
BeXB.

Observed by TESS in sectors 18 (November 2019), 24 (April 2020), and 25
(May 2020), the light curves display a modulation whose amplitude varies in
time. The periodogram shows three groups of frequencies at 0.13, 0.67, and 1.4
\cd. The third group is located at approximately twice the frequency of the
second group and displays smaller amplitudes. The frequency of the most significant 
peak of the second group is 0.67 \cd with an amplitude of 0.9\% in flux (or 9 ppt).  
%Despite its massive and hot optical companion, we did not find signals above 2 \cd,
%typical of $\beta$ Cephei pulsators, which share similar mass, temperature and
%luminosities. 
%The group pattern and low-frequency modulations favor a rotational
%origin of the signals. 
The frequency analysis detects up to 62 peaks with
$S/N>4$. However, after removing all possible combinations of frequencies and
harmonics as described in Sect.~\ref{indfreq}, only four frequencies remained as
independent. These are the three frequencies mentioned above plus another one at
0.75 \cd. \\

{\it RX\,J0146.9+6121/LS I +61 235} \\

LS\,I+61\,235 (V\,831 Cas) is the optical counterpart of the BeXB
RX\,J0146.9+6121 \citep{reig97b}. It is a B1Ve star located in the open cluster
NGC\,663. The binary contains a slowly rotating ($P_{\rm spin}=1412$ s) neutron
star on a wide but unknown orbit \citep{motch97}.  \citet{sarty09} detected
three distinct periodicities that were interpreted as pulsation in the radial
fundamental mode (2.9 \cd), the spin of the Be star (1.5 \cd), and a higher
order $p$-mode pulsation (9.7 \cd). 

 We observed LS\,I+61\,235
extensively for four consecutive nights from Skinakas in September 2018 with about 600
measurements per night in the $V$ and $B$ bands (Table~\ref{logobs1}). We also analyzed two older runs
performed ten years earlier. In our analysis of the
2018 data, we detect three significant frequencies, two of which coincide with
those reported by \citet{sarty09}, namely, 2.9 \cd and 9.70 \cd, and the
third one, at 1.36 \cd, only differs slightly. However, whereas the
higher-frequency modulations can be considered as firm detection, the 1.36 \cd
frequency has low significance with $S/N=3.7$. The two significant frequencies
at 9.7 \cd and 2.9 \cd correspond to periods of 2.5 hr and 8.3 hr, which are too short to be associated with the rotational period of the star. 
%They are consistent with the type of pulsations seen in $\beta$ Cep stars.
%\citep{stankov05,burssens19}. 
There are no TESS observations for this source.\\

%\subsection{IGR\,J010583+6713}

%The X-ray transient IGR\,J01583+6713 was discovered by the IBIS/ISGRI imager
%onboard {\it INTEGRAL} during an observation of the Cas A region on 2005
%December 6 \citep{steiner05}. The compact companion is a strongly magnetized
%neutron star, but pulsations have not yet been firmly established \citep{wang10}.
%\citet{kaur08} identified the spectral type of the companion star to be B2 IVe.

{\it RX\,J0240.4+6112/LS\,I +61 303} (Fig.~\ref{lc1})\\

RX\,J0240.4+6112 is an intriguing BeXB that has been detected over all the electromagnetic
spectrum, from radio to $\gamma$-rays
\citep{taylor92,paredes97,albert06,grundstrom07c,zamanov13}. The emission is
modulated with a period of 26.5 days, which is interpreted as the orbital period
\citep{aragona09}. The nature of the compact companion is not known, although
the lack of pulsation may indicate the presence of a black hole. The optical
counterpart is the B0Ve star LS\,I +61 303.

LS\,I +61 303 was observed by TESS in sector 18 (November 2019). In addition to the short-term variability (hours to days), the light
curve shows significant intensity changes on timescales of weeks. The periodogram is dominated
by red-noise below 2 \cd. The frequency analysis displays the two frequency
groups pattern, with groups centered around 0.9 and 1.8 \cd, together with
several other significant peaks at lower frequencies between 0.15 and 0.55 \cd.
No significant peaks at frequencies higher than 3 \cd were detected. \\

{\it Swift\,J0243.6+6124} (Fig.~\ref{lc1})\\

Swift\,J0243.6+6124 is a unique system. It is the first and only ultra-luminous
X-ray source in our Galaxy \citep{wilson18}. It is the first and only high-mass
X-ray pulsar showing a radio jet \citep{vandeneijnden18}. It was first detected
by Swift/BAT on 3rd October 2017 \citep{kennea17} and confirmed as a BeXBs two
days later \citep{kouroubatzakis17}. The binary harbors a pulsar with a spin
period of 9.9 s  \citep{wilson18} and a $V=12.9$, O9.5Ve star, located at a
distance of $\sim$5 kpc \citep{reig20}.

Our ground-based observations from October 2019 did not reveal any significant
frequency in the interval 0 to 20 cd$^{-1}$. The accuracy of the differential
photometry is estimated to be 7 mmag, in both the $B$ and $V$ bands.

Swift\,J0243.6+6124 was observed by TESS during sector 18 (November 2019)
only. Non-periodic variability seems to be the most characteristic feature of
the periodogram. The low-frequency part of the periodogram is populated by
numerous low-amplitude, low significance peaks.  At first glance, there
seems to be two groups of frequencies, not harmonically related at around 0.2
and 1.1 \cd. The most significant peak of the first group appears at  0.22 \cd
(4.8 days) and manifests as an irregular sinusoidal modulation over the entire
duration of the observations with an amplitude of 0.14\% in flux. Less
significant is the peak at 0.87 \cd of the second group with less than 0.1\%
amplitude. Above 3 \cd, the frequency spectrum is consistent with stochastic
noise, with the exception of two barely significant frequencies at 3.2 and 3.7
\cd. \\

%The 0.22 \cd modulation can be interpreted as a $g$-mode pulsation. \\

%frequencies at around $\sim0.2-1$ \cd $\sim 1.0-1.3$ \cd are detected. Some of
%the lower frequencies can be interpreted as  the beating frequencies of the
%higher frequency group. 

%The $\sim 1$ \cd frequencies are typical of SPB, although a rotational origin
%cannot be ruled out.

{\it V\,0332+53} (Fig.~\ref{lc1})\\

The bright X-ray transient source V 0332+53 was discovered by the Vela 5B
satellite during an outburst in 1973 \citep{terrell84}. The binary consists of a
neutron star with a spin period of 4.4 s \citep{stella85} and a V=15.3, O8--9 Ve star
\citep{negueruela99}. The orbital period is 33.8 days \citep{doroshenko16}.

V\,0332+53 was observed by TESS in sectors 18 and 19 (November 2019). The 
periodogram is dominated by two independent frequencies at  1.46 \cd, and 2.38
\cd and two closely spaced frequencies at 0.27 and 0.31 \cd. Another peak at 3
\cd, most likely the harmonic of the 1.5 peak,  stands above the local noise
level only when we restrict the noise calculation to the close vicinity of the
peak. These peaks lie on top of three groups of frequencies, although they are
not harmonically related. 
%The frequencies of the peaks are typical of SPB pulsators. 
%However, the spectral type of the optical
%companion to V\,0332+53, an O8--9Ve star, place it in the region of $\beta$
%Cephei pulsators, where $p$-modes with frequencies above $\sim3$ \cd are expected. 
TESS data does not detect any modulation above the noise level
at high frequencies.   The red noise below 3 \cd could be attributed to
inhomogeneities in the disk. 

An alternative explanation for the 1.45 \cd is that it might be related to
rotation if the massive companion is rotating close to the break-up velocity,
which for a late O-type star ($M\sim 22-23 \msun$) would be $\sim600$ km s$^{-1}$. The 1.4 \cd
frequency would roughly correspond to a rotational velocity of that magnitude if
$R_*=8-9\rsun$. The low inclination angle of this system
(10$^{\circ}$--20$^{\circ}$) and a projected rotational velocity of $v\sin
i\approx 150$ km s$^{-1}$ are not at odds with such possibility
\citep{negueruela99,zhang05,caballero16}. \\

{\it X\,Per}  (Fig.~\ref{lc1})\\

X Persei (HD 24534) was detected in X-rays with the {\it Uhuru} satellite and
identified as the optical counterpart of the  X-ray source 2U 0352+30
\citep{Braes72}. The main component of X\,Per is a hot massive rapidly rotating
B0Ve star \citep{fabregat92,lyubimkov97}. The secondary is a slowly spinning neutron star
($P_{\rm spin}= 837$ s).
\citet{delgado01} determined an orbital period of 250 days and orbital
eccentricity of $e=0.11$.

X Persei was observed by TESS in sectors 18 (November 2019), 43 (September
2021) and 44 (October 2021). The periodograms show the characteristic two
frequency groups pattern, with groups at 1.8 and 3.6 \cd. There is a significant
frequency at 2.89 \cd, isolated from any group, which appears in sectors 43 and
44, but is not present in sector 18. There are also significant high frequency peaks  with very small amplitude which show up in one of the
sectors, but not in the others  (8.62 \cd, 0.04\% amplitude, sector 18; 4.29
\cd, 0.04\%, sector 44). The light curve presents long term quasi-periodic variability, although, unlike 1A\,0535+26, there are no significant variations in the amplitude of the short term oscillations.
%This might indicate that X per is a hybrid pulsator,
%but due to the small amplitude and the lack of stability of the frequencies in
%the different sectors, this is very uncertain. \\
\\

{\it RX\,J0440.9+4431/LS\,V\,+4417} (Fig.~\ref{lc1})\\

LS V +4417 is a  bright V = 10.8, B0.2Ve star that is associated with the X-ray
source RX J0440.9+4431 \citep{reig99}.  LS V +4417 exhibits long-term optical
photometric and spectroscopic variability linked to the evolution of the
circumstellar disk around the Be star \citep{reig05b,yan16}. Since 1995, the source
has gone through two low-optical states and two high-optical states. The
low-optical states correspond to the (almost) disappearance of the disk.  In the X-rays, it showed one major
outburst in 2010 followed by two less intense flares, which allowed the
determination of the orbital period in 150 days \citep{ferrigno13}.

We observed LS\,V\,+4417 in January 2019. The observations took place from the Aras
de Los Olmos observatory.  The light curve consistently showed a periodicity at
3.20 \cd, regardless of the comparison stars used. However, in all cases the
signal-to-noise was too low ($S/N=3.5$). In favor of the reality of the modulation is the fact that the analysis of
different combination of comparison stars without including the object did not
yield this frequency. Therefore, we conclude that it is very likely associated with the
BeXB.

TESS observed LS\,V\,+4417 during sector 19 (December 2019).  The
periodogram below 3 \cd  is populated by several narrow peaks on top of frequency
groups. The most significant signals are at 2.60 \cd, 2.17 \cd, and
1.10 \cd  with amplitudes of 0.8\%, 0.6\%, and 0.3\%, respectively. A strong isolated peak at 6.74 \cd dominates
the periodogram above 3 \cd.  
\\

{\it 1A\,0535+262} (Fig.~\ref{lc1})\\

1A\,0535+262 is one of the first BeXBs to be discovered \citep{rosenberg75}. Its
optical companion is the V=9 mag, B0IIIe star V725 Tau
\citep{rappaport76,janot87}. The system contains an X-ray pulsar with a pulse
period of 103 s, which orbits the Be star in a moderately eccentric ($e =
0.47\pm0.02$) orbit with an orbital period $P_{\rm orb} = 111.1\pm0.3$ d
\citep{finger96}.  

1A\,0535+262 was observed from the Skinakas Observatory for six days in October
2009. No significant frequency was detected in the periodogram.  
The star was observed in sectors 43 to 45 by TESS (September-November 2021). The light curve shows
long term quasi-periodic variability and also short-term variability. In the
normalized light curve, the short period variability is apparent, with variable
amplitude most likely produced by the beating of several modes of closely spaced
frequencies. 
The long term and short period variability are correlated, in the sense that the maximum amplitude of the short term oscillations coincides with the brightening of the star immediately after each minima, while the minimum amplitude takes place at the fading after each maxima (see Sect.~\ref{slow}).  The frequency analysis performed on the TESS light curve
revealed the characteristic pattern of two frequency groups, centred around 1
and 2 \cd. Amplitudes in the latter are much larger than in the former. The most
significant frequency is at 2.13 \cd, with an amplitude of about 1\% in flux.
There are also two other groups of significant frequencies around 3 and 4 \cd
with much smaller amplitudes. In addition, there are isolated significant
frequencies detected at 6.10 and 7.18 \cd, with amplitudes of 0.03\%. %These frequencies are characteristic of $\beta$ Cep pulsators, and hence 
%1A\,0535+26 appears to be an hybrid SPB/$\beta$ Cep pulsator. 
\\

\begin{figure*}
   \centering
   \includegraphics[width=17cm,clip]{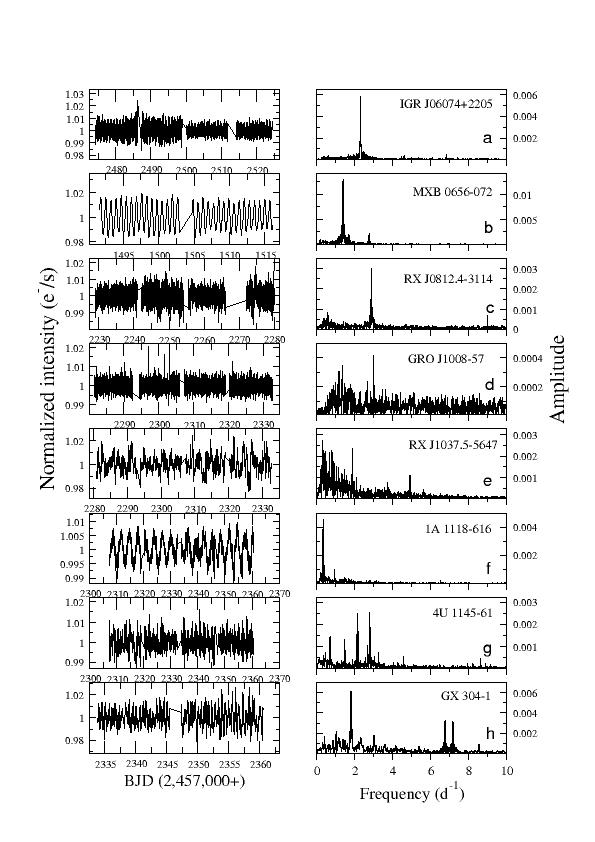}
      \caption{Light curves and periodograms. The time in the light curves is the Barycentric Julian Date with reference time 2,457,000.}
	 \label{lc2}
   \end{figure*}

{\it IGR\,J06074+2205} (Fig.~\ref{lc2})\\

IGR\,J06074+2205 was discovered by {\it INTEGRAL}/JEM-X in February 2003
\citep{chenevez04}.  \citet{halpern05} suggested that a Be star located within
1$'$ of the {\it INTEGRAL} position was the optical counterpart.
\citet{reig10b} confirmed the identification and performed a detail study of the optical emission from IGR\,J06074+2205.
The primary companion of the binary is a $V = 12.3$, B0.5Ve star located at a distance of $\sim$4.5
kpc, showing spectral variability both in the strength and shape of the \ha\
line. IGR\,J06074+2205 is an X-ray pulsar with a spin period of 373.2 s
\citep{reig18a}.

$R$ and $V$-band photometric observations were carried out in February 2019 from
the Aras de los Olmos observatory. We did not detect any significant modulation.

%We detected a periodicity at 3.33 \cd, albeit in the limit of what it can be
%considered as a firm detection, $S/N=4.0$.

%The TESS periodogram is similar to that of 1A\,0535+262, with a very
%strong modulation at 2.3 \cd. After pre-whitenning for this frequency, the
%periodogram is dominated by stochastic variability, characterized by red noise
%and 3 modulations below 0.1 \cd.  This source also shows a low-amplitude $\beta$
%Cephei pulsation at 6.8 \cd. The red noise and low frequencies may reflect
%changes in the disk, while the well defined and coherent oscillations can be
%attributed to NRP of $\beta$ Cephei type.

Observed in sectors 43 and 44, the TESS periodogram is characterized by
the absence of frequency groups and a highly significant peak dominating the
0--20 \cd spectrum. The main feature is located at 2.30 \cd, with an amplitude around 0.5\% in flux and $S/N>45$. Its first harmonic  at 4.60 \cd is also apparent. A second frequency at 2.15
\cd is also significant, albeit with a much smaller amplitude. Another significant frequency is found at 6.83 \cd.
%which would indicate that the star is a hybrid SPB/$\beta$ Cep pulsator. 

We note that the  photometric aperture includes
light from the nearby source  Gaia\,DR3\,3423526544838561792, located 26.2$\arcsec$ away. This contaminating star has a Gaia magnitude G=12.80, to be compared with G=12.17 for IGR\,J06074+2205. Their astrophysical parameters given by Gaia DR3, $T_{\rm eff}=6347$ K, $\log g = 3.530$, correspond to a F6III spectral type \citep{gray05}.
\\

{\it MXB\,0656--072} (Fig.~\ref{lc2})\\

The transient X-ray source MXB\,0656--072 was discovered on 1975 September 20
with {\it SAS-3} \citep{clark75}, and identified with a 160-s X-ray pulsar by
\citet{remillard03}. Based on the periodic variation of the X-ray intensity
during the active phases, an orbital period of 101.2 days was proposed by \citet{yan12a}.
However, an orbital solution of the system remains undetermined. \citet[][see
also \citealt{nespoli12}]{pakull03} identified the optical counterpart with an
O9.7 Ve star, hence MXB\,0656--072 was classified as a BeXB.

TESS observed the source in sector 7 (January 2019). At first glance, the frequency spectrum seems to exhibit three frequency groups at around 0.2, 1.4 and 2.8 \cd.  However, the grouping of frequencies resembles that of the window function rather than the characteristic structure seen in other BeXBs. Indeed, the third group (the harmonic of the 1.38 \cd peak) is formed by the central frequency and two side lobes. The main frequency at 1.38 \cd ($S/N=100$) has an amplitude of 1.3\% in flux, the largest of all the BeXBs. The 1.38 \cd  modulation  corresponds to a period of 0.71 days. If interpreted as the rotation of the Be star, then the star must be rotating very close to the critical velocity, which for an O9.7V star would be $v_{\rm crit}\approx 550-560$ km s$^{-1}$ \citep{townsend05}. A frequency of  1.4 \cd implies a rotational velocity of  $520-560$ km s$^{-1}$ assuming a radius of $7.5-8 \, \rsun$. \\

{\it RX\,J0812.4--3114/LS\,992} (Fig.~\ref{lc2})\\

RX\,J0812.4--3114 was first identified by the ROSAT Galactic Plane Survey as an
X-ray source that positionally coincides with the B0.2IVe star LS 992
\citep{motch97,reig01}. It harbors an X-ray pulsar with spin period of 31.9
seconds \citep{reig99b}. This source spends most of its life in a very low X-ray
state. Occasionally it enters an active period where a series of weak X-ray
outbursts are detected. Based on the regularity of these outbursts,
\citet{corbet00} estimated the orbital period to be 80 days.

RX\,J0812.4--3114 was observed in sectors 8 and 9 (February-March 2019) and 34 and 35
(January-February 2021). In sectors 8 and 9 a three frequency groups pattern is
apparent, with groups centered at 0.3, 1.3 and 2.6 \cd. The two main peaks of the harmonically
related (higher frequency) groups have amplitudes of 0.3\% and 0.5\% respectively.
In addition, there is another significant frequency well detached from the
second group at 2.89 \cd, with amplitude of 0.5\%. In sectors 34 and 35 the frequency spectrum changed
significantly. The frequency groups pattern vanished, while the isolated
frequency at 2.89 \cd is still present, with an slightly smaller
amplitude. Moreover, in these two last sectors, two high frequencies at 9.00
and 11.89 \cd emerged, which are not present in sectors 8 and 9. \\
%If we interpret
%the frequency groups as produced by rotational modulation, their disappearance
%in sectors 34 and 35 would imply that the photospheric region of the star is
%more homogeneous than in sectors 8 and 9. This new photospheric structure would
%be related with the visibility of the high frequency modes at 9.00
%and 11.89 \cd. \\

{\it Vela X--1} (Fig.~\ref{lc3})\\

Vela X--1 is the archetype of a classical supergiant X-ray binary where a 283-s
pulsating neutron star is on an approximate nine-day eclipsing orbit around HD
77581, a B0.5 Ib star \citep{mcclintock76}. The hot primary loses mass at a rate
of $\sim10^{-6} \msun {\rm yr^{-1}}$  \citep{kretschmar21} through a powerful stellar wind, which is
accreted onto the neutron star and powers the X-ray pulsar. The neutron star is
eclipsed by the primary during every orbital cycle. The mass and radius of the
primary are estimated to be  $\sim23\,\msun$  and  $\sim30\,\rsun$, respectively
 \citep{vankerkwijk95,kretschmar21}.

Vela X--1 was observed by  TESS in sectors 8 and 9 (February-March 2019). The
periodogram is very similar to that of 2S\,0114+65. It is completely featureless
above 2 \cd, while at lower frequencies there are a myriad of peaks, whose
amplitudes decrease with frequency (i.e. red noise). A strong peak at  0.22 \cd
dominates the low-frequency part. The amplitude of this peak is 3\% in flux,
which is the highest amplitude measured in all our sample of sources. \\

{\it GRO\,1008--57} (Fig.~\ref{lc2})\\

This BeXB was discovered by the Burst and Transient Source Experiment (BATSE) on board
the Compton Gamma-Ray Observatory ({\it CGRO}) during a 1.4 Crab giant outburst
in 1993 \citep{stollberg93}.  Optical follow up observations identified an
early-type star with infrared excess and strong H$\alpha$ emission as its Be-type
companion and suggested a distance to the source of 5 kpc \citep{coe94,coe07}.
The compact companion is a neutron star with a spin period of 93.5 s. The binary
has an orbital period of 249.48 days and an eccentricity of 0.68
\citep{coe07,kuhnel13}. The magnetic field of the pulsar is known to be the
highest among BeXBs, likely as high as $8\times10^{12}$ G, as measured from the suggested
cyclotron line at 88 keV \citep{shrader99,bellm14}. GRO\,J1008--57 shows both
giant irregular type II outbursts and regular type I outbursts \citep{kuhnel17}.

TESS observed GRO\,1008--57 in sectors 9 and 10 (March-April 2019) and  36
and 37 (March-April 2021) during the first and third year of the mission,
respectively. The frequency spectrum at high frequencies is similar in both
epochs and it is dominated by two isolated peaks at 2.98 and 10.2 \cd. The periodogram
differs at low frequencies. During the first epoch, the power concentrates mainly
at $\simless1$\cd, while during the 2021 observations the power is mainly shared in
the range $0.5-2$ \cd.  The overall shape of the periodogram is reminiscent of a
ROTD pattern \citep{balona17}, consisting of a broad peak flanked by a
sharp peak at slightly higher frequency. The mechanism that produce such a
pattern is unknown, although \citet{balona17} proposed that the broad peak could
be associated with starspots, while the sharp peak with an orbiting body. A more likely explanation is given by \citet{saio18}, who identify the sharp peak with the rotation frequency and the broad hump as even r modes (Rossby waves).\\

{\it RX\,J1037.5--5647/LS\,1698} (Fig.~\ref{lc2})\\

This source was first detected by {\it Uhuru} \citep{forman78}, but it is also
associated with the {\it Ariel V} source 3A 1036--56 \citep{warwick81}. The system
was detected again during the ROSAT Galactic plane survey and named RX\,J1037.5--5647.
{\it RXTE} data revealed pulsed emission with a period of $860\pm2$ s
\citep{reig99}, while a tentative 61 day orbital period has been proposed from
the analysis of {\it Swift} BAT and XRT data \citep{cusumano13}. Based on a
pointed ROSAT Position Sensitive Proportional Counter (PSPC) observation,
\citet{motch97} identified the optical counterpart to be the B0 III-Ve star LS
1698, at a distance of $\sim$5 kpc.

Observed in sectors 10 (April 2019), 36 (March 2021), and 37 (April 2021), the
star does not present the frequency groups pattern. Instead, the frequency spectrum is
dominated by low-frequency noise with a strong red-noise component. The most
significant frequencies are at 0.31 and 1.86 \cd. The frequency at 3.72 \cd seems to be the first harmonic of the
1.86 \cd frequency. There are two stable high frequencies at 4.91 \cd and 5.64 \cd. Several short-lived significant frequencies are also apparent between 0.2 and 1.2 \cd and between 3 and 4 \cd, but most of them appear either in one observing period or in the other, but not in both.  \\

{\it 1A\,1118--616} (Fig.~\ref{lc2})\\

The X-ray transient 1A\,1118--616 was discovered during an outburst in 1974 by
the Ariel-5 satellite \citep{eyles75}.  The optical counterpart was identified
as the Be star He 3-640/Wray 793, with an O9.5IV-Ve spectral type.
\citep{janot81}. The system contains a slowly rotating neutron star with a
spin period of 406.5 s \citep{ives75}. Since its discovery, 1A\,1118--616 has
displayed three giant X-ray outbursts, in 1974, 1992, and 2004 \citep{nespoli11}. The Be star in
this system displays the strongest \ha\ line in emission among BeXBs, with an
equivalent width of $EW(H\alpha)\sim -90$ \AA\ \citep{coe94}. One peculiarity of this BeXB is its unusual short orbital
period.  According to the $P_{\rm orb}-P_{\rm spin}$ correlation that holds for
BeXBs \citep{corbet86}, the orbital period should be longer than $\sim200$ days.
However, \citet{staubert11} reported an orbital period of only 24 days.

1A\,1118--616 was observed by TESS in sectors 10 and 11 (March-May 2019) and
37 and 38 (April-May 2021). The variability pattern between the
two TESS epochs changed significantly. The 2019 light curve is more erratic and displays low amplitude fast variability, whereas the 2021 light curve shows a distinct
sinusoidal variation with a period of $\sim$3 days. In sectors 37 and 38, a visual inspection of the frequency spectrum reveals a three frequency groups pattern, at frequencies of 0.3, 1.6, and 3.1 \cd. In sectors 10 and 11, the two first groups have almost merged in a broad, single group.
The light curve in sectors 37 and 38
is dominated by a 0.33 \cd frequency with an amplitude of 0.5\% in flux.  This frequency is also present in sectors
10 and 11, although with a much smaller amplitude, 0.15\% in sector 10 and less
than 0.05\% in sector 11. There is also a strong peak at 0.94 \cd which appears
in all sectors with amplitudes between 0.1 and 0.2\%, and a frequency of 1.16 \cd
with similar amplitudes present only in sectors 10 and 11.\\ 

%All these frequencies
%are characteristic of SPB-type pulsations.  
%There are also significant frequencies at 4.42, 6.53 and %14.01 \cd, which are
%characteristic of $\beta$ Cep type pulsations. The star is hence a hybrid
%pulsator. \\

{\it 4U\,1145--61} (Fig.~\ref{lc2})\\

The X-ray pulsar 4U\,1145-619 was discovered in 1972 by {\it Uhuru}
\citep{giacconi72}. The optical counterpart of 4U\,1145-61 is the V=9 mag, B1 Ve
V801 Cen \citep{stevens97,alfonso17}. X-ray pulsations with $P_{\rm spin}=292$ s were
detected in 1977 with {\it Ariel V} \citep{white78}. Analysis of the long-term
X-ray behavior of 4U\,1145--61 revealed recurrent outbursts with a period of
186.5 d. Outbursts are typically of 10-d duration, with flux levels increasing
by a factor of $\sim5$ \citep{watson81,priedhorsky83,warwick85,nakajima15}.

TESS observed 4U\,1145--61 during the first year (sectors 10 and 11,
April-May 2019) and third year (sectors 37 and 38, two years later) of the
mission.  This source shows a densely populated frequency spectrum above 3 \cd,
with at least three groups of frequencies  at around 4.2 \cd, 8.4 \cd, and 12.8 \cd and
some smaller amplitude modulations at 6.5 \cd and 10.2 \cd.  These groups are
visible during the two epochs, but the 12.8 and 10.2 \cd signals have $S/N< 4$ in sectors 10 and 11. At the lower end of the periodogram, there are a 
number of narrow peaks with an amplitude about 4 or 5 times larger than the
amplitude of the high-frequency groups and $S/N\simmore 10$.\\  

%The frequencies of these peaks are 0.38
%(sector 10), 0.69 (sectors 37 and 38), 1.46--1.50, 2.10, 2.16, 2.66 (sector 38),and 2.80 \cd. 

%The 0.7 \cd and 1.5 \cd could be attributed to rotation and the corresponding
%harmonic, whereas the peaks in the range 2--3 \cd lie in between those typical
%of SPB and $\beta$ Cephei stars. 

%The higher frequencies peaks are different from
%the typical $p$-mode pulsations seen in $\beta$ Cep stars. %The fact that they
%appear as multiplet structures (rather than as isolated peaks) and that they are
%harmonically related suggest a rotational origin \citep[see e.g.][]{papics12}. \\

{\it GX\,304--1/4U\,1258-61} (Fig.~\ref{lc2})\\

This BeXB was first detected in X-rays during balloon observations in October
1967 \citep{lewin68} and recognized as an X-ray pulsar, with $P_{\rm spin}=272$ s, by
\citet{mcClintock77}.   The X-ray long-term variability of GX\,304--1 is
characterized by periods of strong activity where type I outbursts recur
periodically with a period of 132.2 days, which is considered to be the orbital
period \citep{priedhorsky83,sugizaki15}, and periods of hardly any activity. The optical counterpart is a B2Vne star \citep{parkes80}.

Observed in sector 38 (April-May 2021) by TESS, the periodogram does not
show the frequency groups pattern. There are several significant, isolated
frequencies detected from 1.02 up to 15.33 \cd. The larger amplitudes are found at 1.80 \cd (0.6\% in flux),  6.77 \cd (0.3\%), and 7.18 \cd (0.3\%).  The 1.80 \cd could also been
attributed to rotation if the star is rotating close to the critical velocity.
Assuming a radius of $\sim 5\,\msun$ for a spectral type B2V, the 1.8 \cd
frequency would correspond to a rotational velocity of 450 km s$^{-1}$, while
the critical velocity is estimated to be 475 km s$^{-1}$ \citep{townsend04}. The
fast-rotating nature of this star was already noticed by \citet{parkes80}, who
suggested that it is rotating at near break-up velocity. \\

{\it KS\,1947+300} (Fig.~\ref{lc3})\\

\ks\ is associated with a moderately reddened, V = 14.2, B0Ve star, located at
$\sim$10 kpc \citep{negueruela03}. Its optical and infrared emission is
rather stable with minor optical outbursts of 0.1 mag and correlated X-ray
activity \citep{kiziloglu07a}.

Ground-based observations of the optical counterpart to \ks\ were performed from the Skinakas observatory during three consecutive nights on
28-30 July 2010 and on 4-6 August 2011 (Table~\ref{logobs2}). The analysis of the
2010 light curve does not yield any significant frequency. The highest peak is
found at 5.41 \cd with an amplitude of 2 mmag and very low signal-to-noise ratio
$S/N=3.2$. 
The 2011 light curve contains two low-amplitude but significant peaks at 2.31 \cd and
amplitude 3.9 mmag ($S/N=12$) and  2.73 \cd with amplitude 2.2 mmag ($S/N=6$). 

The source was observed by TESS in sector 41 (July-August 2021). The
periodogram is dominated by noise, with only two significant peaks at 0.5 and 3.6
\cd. Another peak at the expected frequency of the first harmonic of the 3.6
\cd modulation stands above the local noise level, although it does not meet the
$S/N>4$ criterion. \\

{\it Swift\,J2000.6+3210} \\

Swift\,J2000.6+3210 is a scarcely observed BeXB discovered by the {\it Swift}/BAT
instrument in 2005 \citep{tueller05}. Subsequent optical observations reported
\ha\ emission with $EW(H\alpha)=-10$ \AA\ \citep{burenin06,halpern06,masetti08},
placing the system as a strong BeXB candidate. The most complete X-ray analysis
was performed by \citet{pradhan13}, who detected pulsations with a period of 890 s.

We observed Swift\,J2000.6+3210 from Skinakas during 28-31 July 2021 and 19-22 August
2021 (Table~\ref{logobs2}). The timing analysis gave a significant ($S/N=10$) peak at 2.8 \cd\
during the July run. The weather condition during the August observations were
not as good as in July, resulting in a noisier light curve. The frequency at 2.8
\cd\ is detected with low significance ($S/N=4$). No TESS light curve of this source has been analysed, because even the smallest one pixel photometric aperture is heavily contaminated from a nearby bright star. \\

{\it GRO\,J2058+40} (Fig.~\ref{lc3})\\

GRO\,J2058+40 is a BeXB with an O9.5-B0 companion at about 9 kpc 
\citep{reig05a,wilson05}. It was discovered by BATSE on board the
Compton Gamma-Ray Observatory (CGRO) during its giant
outburst in 1995 September-October \citep{wilson98}. The compact companion is a
198-s pulsar in a 55 day orbit. In the optical band, GRO\,J2058+40 is a
long-term variable system with photometric and spectroscopic changes on time
scales of years. These changes are attributed to the growth and dissipation of a
circumstellar disk \citep{reig15,reig16}. In addition, two modulations at 2.37
\cd and 2.40 \cd were reported in the Robotic Optical Transient Experiment1 (ROTSEIIId)
light curve  and attributed to non radial pulsations \citep{kiziloglu07b}.

We observed GRO\,J2058+40 for four consecutive nights in July 2020 from Skinakas
and detected the 2.40 \cd frequency again (Table~\ref{logobs2}). In addition, we detected two other
frequencies at 2.64 \cd and 4 \cd.

TESS observed this source in sector 15 (August 2019). Two isolated peaks
are detected at 2.37 and 6.38 \cd. The lower-frequency peak is the strongest
signal in the periodogram and coincides with the one detected in ground-based
observations. 
%These frequencies are characteristic of SPB and $\beta$ Cep
%pulsations respectively, and hence the star is a hybrid pulsator. 
There is some
evidence of frequency groups at 2.0 and 4.0 \cd, although somewhat uncertain. \\

{\it SAX\,J2103.5+4545} \\

\sax\ is the BeXB with the shortest known orbital period, $P_{\rm orb}=12.7$ d \citep{baykal00,camero07}.
Due to the short orbital period and moderate eccentricity, the neutron star
truncates the Be star's disk at a small radius and prevents the development of
an extended and steady disk. This translates into fast spectral changes and
asymmetric H$\alpha$ line profiles \citep{reig10a,kiziloglu09,camero14,reig16}.
The primary is a moderately reddened ($A_V=4.2$) early B-type (B0Ve) star
\citep{reig04}.

The system was observed from the Skinakas observatory in September and October
2009 and July 2020 (Table~\ref{logobs2}). The 2009 light curve contains 262
data points, distributed in 13 nights with an average of  a point every 20
minutes. The reason for this sampling is that another source,  \vcas,  was
observed on the same nights. The telescope pointed alternatively to these two
systems. The 2020 observations had a high sampling rate with 300-400 points per
night during four consecutive nights. The periodogram shows a low significant
($S/N=4.8$) peak at 3.2 \cd with an amplitude of 3.8 mmag in 2009 and two
significant low amplitude peaks at 2.7 \cd and 2.4 \cd in 2020. As with the other sources observed from ground-based telescopes,  modulations with frequencies between 2--3 \cd should be treated with caution as these periods are similar to the length of the observing nights.

Although the field around this source was observed by TESS, its light
curve is affected by the emission from two nearby stars of similar brightness.
The light curve is very complex and it is possible that one of the nearby stars
is an eclipsing system. It is not possible to distinguish the different origin
of the variability. Hence, we did not include the TESS analysis of SAX\,J2103.5+4545
in this work. \\

{\it IGR\,J21343+4738} \\

IGR\,J21343+4738 was discovered by the Imager on Board the {\it INTEGRAL} Satellite
(IBIS) as a transient source that alternated period of X-ray activity with
periods of quiescence \citep{bird07,krivonos07,bikmaev08}. \citet{reig14b}
reported the discovery of X-ray pulsations during an XMM-Newton observation.
They obtained a barycentric corrected pulse period of 320.35 s.   The optical
counterpart to IGR\,J21343+4738 is a B1IVe star that exhibits spectral and intensity
variability on time scales of months \citep{reig14a}.  The presence of shell
absorption lines indicates that the line of sight to the star lies nearly
perpendicular to its rotation axis (edge-on system).

We observed IGR\,J21343+4738 for three nights in July 2018 and another three
nights in July 2019 from the Skinakas observatory and on 9 separate runs in
July-September 2019 from the Aras de los Olmos observatory (Table~\ref{logobs2}). We detected two
frequencies with a $S/N$ ratio above 4 during the 2018 campaign, at 3.66 \cd and
2.04 \cd, and three during the 2019 campaign at 2.76 \cd, 4.69 \cd, and 2.08
\cd. No TESS observations are reported for this source because the light
curve is affected by a nearby bright source. \\

{\it Cep X--4} \\

%The reason is that after its initial discovery in 1972, Cep X-4 went into
%outburst every four to five years (1988, 1993, 1997, 2002, 2009, 2014, 2018?)

Cepheus X-4 was discovered in 1972 as a transient X-ray source by the OSO-7
satellite \citep{ulmer73}.  Many of the studies of this source have been
performed in the X-ray band owing to its frequent variability that consists of small
X-ray outbursts every four or five years. During the 1988 outburst, X-ray
pulsations at 66.3 s \citep{koyama91} and a cyclotron line at $\sim$ 30 keV
\citep{mihara91} were detected. The first suggestion on the nature of the
optical counterpart was made by \citet{koyama91}, who argued that Cep X-4 is
most likely a BeXB system with an orbital period longer than 23 days.  The
long-term {\it RXTE}/ASM X-ray light curve revealed a tentative periodicity at
20.85 d, which could be attributed to the orbital period \citep{mcbride07}. The
first detailed optical spectroscopic study on this star was carried out by
\citet{bonnet98}.

We observed Cep X--4 from the Skinakas observatory on three nights from 22--24
August 2020 (Table~\ref{logobs3}). Non-sinusoidal large amplitude variations are seen in the light
curve. The V-band light curve shows changes of up to 0.02 mag from maximum to
minimum. Two significant periodicities were found at 2.07 and 5.37 \cd.
No TESS observations are reported for this source because the light
curve is affected by a nearby bright source. \\

%TESS observed Cep X--4 during its second year of operations (sectors 16
%and 17). Two unrelated groups of frequencies at 0.5--0.7 \cd and 1.8--1.9 \cd
%show up in the frequency spectrum. A third group at twice the frequency of the
%second group is also seen at 3.7 \cd, but it is below a $S/N$ of 4.

{\it 4U\,2206+54} (Fig.~\ref{lc3})\\

4U\,2206+54 is known as an X-ray source since the early times of X-ray astronomy
\citep{giacconi72}. 4U\,2206+54 is a peculiar system. Despite the optical
spectrum displays the \ha\ line in emission with a split profile, which is
typical of Be stars, its X-ray emission does not show outbursts. The neutron star
is fed by the stellar wind from the O9.5V star. 4U\,2206+54 is the only (together
with LS\,5039) permanent wind-fed HMXB with a main-sequence donor \citep{ribo06}.
The optical companion has a high  abundance of He and a wind terminal velocity
of $\sim350$ km s$^{-1}$, abnormally slow for its spectral type  \citep{blay06}.
4U 2206+54 is also one of the slowest X-ray pulsars known to date with a spin
period of 5560 s \citep{reig09,finger10,torrejon18}.

4U\,2206+54 was observed from the Skinakas observatory for two nights in October 2008 with high-cadence and
for 13 nights in September-October 2009 with lower cadence (Table~\ref{logobs3}). Intra-night variability at 0.01 mag is clearly detected.
\citet{bugno09} reported a tentative modulation in the light curve of
4U\,2206+54 with 2.6 \cd\ and a $S/N$ of 5.5. We confirm the presence of this
periodicity from our ground-based photometry. During the 2008 observations, we
detected a peak at 2.63 \cd with an amplitude of 5.3 mmag. After removing this
frequency, another peak at 4.7 \cd and amplitude 3.2 mmag dominated the
frequency spectrum.  During the 2009 observations, the 2.6 \cd frequency was
present but less significant ($S/N=4$) than a periodicity at 2.12 \cd
($S/N=6.4$).

4U\,2206+54 was observed by TESS during sectors 16 and 17. The light curve
displays low amplitude erratic variability, more typical of wind-fed systems.
The TESS periodogram shows red noise and stochastic variability,
characterized by a large number of peaks in the frequency range 0.2--2 \cd,
without a clear dominant frequency. Only a peak at 1 \cd appears in common in
both sectors. When the two light curves are combined another significant peak at
0.6 \cd stands above the local noise level. We note, however, that  if the noise
level is computed over a range of 1 \cd (instead of 5 \cd), then no significant
peak is found. Above 3 \cd, it shows a significant
isolated peak at 6.14 \cd. \\

{\it SAX\,J2239.3+6116} \\

SAX\,J2239.3+6116 was discovered by the BeppoSAX wide-field camera as a transient X-ray source during observations of
the supernova remnant Cas A \citep{intzand00}. Subsequent {\it RXTE}/PCA
and {\it BeppoSAX}/MECS-LECS observations during a predicted outburst in July 2001 revealed X-ray pulsations with a pulse
period of $1247.2\pm0.7$ s \citep{intzand01}.
The optical counterpart is a $V = 14.8$ B0Ve star located
at a distance of $\sim$4.9 kpc \citep{intzand00}. 
The
long-term optical variability is characterized by
the slow dissipation of the circumstellar disk around the Be star
companion \citep{reig17}.

SAX\,J2239.3+6116 was observed during four contiguous nights from the Skinakas
observatory on 22--25 August 2011 and 20-23 August 2019 (Table~\ref{logobs3}). It shows the most clear
and distinct modulation of all the sources observed with ground-based telescopes in this work. A periodicity
with frequency 1.97 \cd and amplitude 13 mmag is clearly detected in the 2011
light curve.  Surprisingly, we did not detect any
modulation during the 2019 observations.

SAX\,J2239.3+6116 was observed in sectors 16 (September 2019), 17 (October
2019), and 24 (April 2020). The light curve is quite variable even within the
same sector. For this reason, we performed a frequency analysis in each orbit
separately as well as by sector. In sectors 16 and 17, the periodogram shows three frequency groups at 0.3,
0.8 and 1.7 \cd. There are two other frequencies, with larger
amplitudes, at 0.73 and 1.93 \cd, which are close to the groups
but not in the middle of them. The 0.73 \cd appears in all
orbits as one of the most significant modulations,
while the 1.93 is significant only in the two orbits of sector 24.
The 1.93 \cd frequency can be identified with the one detected from ground based
observations in Skinakas in 2011. As mentioned
above,  Skinakas detected this modulation in 2011
but not in 2019. Likewise, TESS did not detect the 1.93 \cd frequency
during the 2019 observations. The time elapsed  between sectors 16-17 and sector
24 is about 6--7 months. A high frequency peak is detected at 11.2 \cd, although the significance is low (Table~\ref{freq-tess5}).\\

{\it MWC\,656} \\

MWC\,656 is the only known BeXB which likely host a black hole companion.  The
binary nature of MWC\,0656  was first suggested by a 60.37 d periodicity in the
optical light curves \citep{williams10}, and later confirmed through radial
velocity studies \citep{casares12,casares14}. It is probably associated with the
$\gamma$-ray source AGL\,J2241+4454 \citep{munar16}. However, \citet{rivinius22}
revisited the optical spectral variability properties of this system and
concluded that they match with a hot subdwarf companion. In this case MWC\,656
would be a Be+sdOB binary system \citep{wang21}.

We observed MWC\,656 in $V$ and $B$ bands from the Aras de los Olmos observatory
for five nights in October 2019 (Table~\ref{logobs3}). More than $\sim$200 measurements were taken
each night. Large amplitude variations are seen in the individual light curves
with changes from minimum to maximum of $\simmore$ 0.05 mag. Our frequency
analysis resulted in the detection of two frequencies at 2.79 \cd with an
amplitude of 10.8 mmag and at 1.98 \cd with an amplitude of 7.9 mmag.

Observed in sector 16 by TESS, the periodogram displays two groups of
frequencies. However, the main frequencies of the two groups are not in a 2:1
relationship. The second group is clearly dominated by a peak at 1.8 \cd, while
the first group contains a strong peak at 0.8 \cd. Moreover, the first group
includes only three peaks and two the second group, instead of the bunch of
closely spaced frequencies seen in many Be stars. If we restrict the analysis at
frequencies above 3 \cd, then we detect two other isolated significant
frequencies at 3.69 \cd and 4.30 \cd. 
The former might, however,  be the harmonic of the 1.8 \cd. The 1.8 \cd (0.56 days) has been associated with the
rotation of the Be star \citep{zamanov21}.

\section{TESS light curves and periodograms}
\label{plots}

The following figures give the light curves (left panels) and frequency spectra (right panels) of every BeXB for which TESS data could be retrieved.

%-------------------------------------------------------------
%-------------------------------------------------------------

%-------------------------------------------------------------
%-------------------------------------------------------------

%-------------------------------------------------------------
%-------------------------------------------------------------

\section{Significant independent frequencies}
\label{tables-indfreq}

The following tables give the independent frequencies of the BeXB for which TESS data could be retrieved. The frequency analysis was performed for the sectors indicated in parenthesis next to the name of the source.

\begin{table*}
\caption{Frequencies and periods of BeXBS. The numbers besides the source name indicate the TESS sector.} 
\label{freq-tess1}
\center
\begin{tabular}{ccccccc}
\hline
\hline
Frequency	&$\sigma(f)$	&Period	&$\sigma(P)$	&Amplitude	&$\sigma(A)$	&S/N	\\
(\cd)		&(\cd)		&(days)	&(days)		&(ppt)		&(ppt)		&	\\
\hline
\multicolumn{7}{c}{4U\,0115+63 (18)} \\
\hline
  3.331 &   0.011 &   0.300 &   0.003 &     7.29 &     0.16 &     28.0   \\ 
  0.207 &   0.011 &   4.841 &   0.054 &     2.18 &     0.16 &      8.4   \\ 
  1.691 &   0.011 &   0.591 &   0.007 &     1.56 &     0.16 &      6.0   \\ 
\hline
\multicolumn{7}{c}{4U\,0115+63 (24,25)} \\
\hline
  3.327 &   0.005 &   0.301 &   0.001 &    10.09 &     0.14 &     39.5   \\ 
%  1.625 &   0.005 &   0.616 &   0.003 &     1.98 &     0.15 &      7.8   \\ 
  2.941 &   0.005 &   0.340 &   0.002 &     1.77 &     0.15 &      6.9   \\ 
  2.828 &   0.005 &   0.354 &   0.002 &     1.50 &     0.14 &      5.9   \\ 
  1.411 &   0.005 &   0.709 &   0.003 &     1.23 &     0.14 &      4.8   \\ 
  0.294 &   0.005 &   3.398 &   0.016 &     1.36 &     0.14 &      5.3   \\ 
  3.183 &   0.005 &   0.314 &   0.002 &     1.31 &     0.14 &      5.1   \\ 
\hline
\multicolumn{7}{c}{IGR\,J01363+6610 (18)} \\
\hline
  0.671 &   0.011 &   1.489 &   0.016 &     6.97 &     0.11 &     33.6   \\ 
  1.482 &   0.011 &   0.675 &   0.007 &     5.23 &     0.11 &     25.2   \\ 
\hline
\multicolumn{7}{c}{IGR\,J01363+6610 (24,25)} \\
\hline
  0.671 &   0.005 &   1.490 &   0.007 &     8.61 &     0.06 &     94.7   \\ 
  0.750 &   0.005 &   1.333 &   0.006 &     4.46 &     0.06 &     49.0   \\ 
  0.132 &   0.005 &   7.607 &   0.036 &     2.70 &     0.06 &     29.7   \\ 
  1.444 &   0.005 &   0.692 &   0.003 &     3.02 &     0.06 &     33.2   \\ 
\hline
\multicolumn{7}{c}{RX\,J0240.4+6112 (18)} \\
\hline
  0.253 &   0.010 &   3.955 &   0.041 &     4.28 &     0.12 &     14.5   \\ 
  0.928 &   0.010 &   1.077 &   0.011 &     2.86 &     0.12 &      9.7   \\ 
\hline
\multicolumn{7}{c}{SWIFT\,J0243.6+6124 (18)} \\
\hline
  0.218 &   0.011 &   4.582 &   0.052 &     1.39 &     0.07 &     10.1   \\ 
  0.872 &   0.011 &   1.147 &   0.013 &     0.83 &     0.07 &      6.0   \\ 
  0.075 &   0.011 &  13.321 &   0.151 &     0.74 &     0.08 &      5.3   \\ 
  1.122 &   0.011 &   0.892 &   0.010 &     0.68 &     0.07 &      4.9   \\ 
  3.172 &   0.011 &   0.315 &   0.004 &     0.31 &     0.10 &      5.2   \\ 
  3.731 &   0.011 &   0.268 &   0.003 &     0.31 &     0.10 &      5.2   \\ 
\hline
\multicolumn{7}{c}{V\,0332+53 (18,19)} \\
\hline
  2.385 &   0.005 &   0.419 &   0.002 &     2.30 &     0.11 &     11.1   \\ 
  1.456 &   0.005 &   0.687 &   0.004 &     2.23 &     0.11 &     10.8   \\ 
  0.269 &   0.005 &   3.714 &   0.019 &     1.53 &     0.11 &      7.5   \\ 
  1.333 &   0.005 &   0.750 &   0.004 &     0.96 &     0.11 &      4.7   \\ 
  0.175 &   0.005 &   5.708 &   0.029 &     0.99 &     0.11 &      4.8   \\ 
  1.387 &   0.005 &   0.721 &   0.004 &     0.97 &     0.11 &      4.7   \\ 
  3.000 &   0.005 &   0.333 &   0.002 &     0.55 &     0.15 &      6.4   \\ 
% 19.389 &   0.005 &   0.052 &   0.000 &     0.34 &     0.14 &      4.1   \\ 
\hline
\multicolumn{7}{c}{ X\,Per (18)} \\
\hline
  1.742 &   0.010 &   0.574 &   0.006 &     2.82 &     0.11 &     10.6   \\ 
  3.608 &   0.010 &   0.277 &   0.003 &     1.75 &     0.11 &      6.6   \\ 
  0.200 &   0.010 &   4.999 &   0.051 &     1.39 &     0.12 &      5.2   \\ 
  3.842 &   0.010 &   0.260 &   0.003 &     0.98 &     0.15 &     12.1   \\ 
  3.211 &   0.010 &   0.311 &   0.003 &     0.70 &     0.15 &      8.7   \\ 
  8.625 &   0.010 &   0.116 &   0.001 &     0.35 &     0.15 &      9.1   \\ 
  4.213 &   0.010 &   0.237 &   0.002 &     0.33 &     0.15 &      4.5   \\ 
\hline
\end{tabular}
\end{table*}

\begin{table*}
\caption{Frequencies and periods of BeXBS (cont.).} %The numbers besides the source name indicate the TESS sector.}
\label{freq-tess2}
\center
\begin{tabular}{ccccccc}
\hline
\hline
Frequency	&$\sigma(f)$	&Period	&$\sigma(P)$	&Amplitude	&$\sigma(A)$	&S/N	\\
(\cd)		&(\cd)		&(days)	&(days)		&(ppt)		&(ppt)		&	\\
\hline
\multicolumn{7}{c}{ X\,Per (43,44)} \\
\hline
  3.610 &   0.005 &   0.277 &   0.001 &     2.80 &     0.05 &     14.5   \\ 
  1.738 &   0.005 &   0.575 &   0.003 &     2.87 &     0.05 &     14.8   \\ 
  2.885 &   0.005 &   0.347 &   0.002 &     1.96 &     0.05 &     10.1   \\ 
  1.838 &   0.005 &   0.544 &   0.003 &     1.86 &     0.05 &      9.6   \\ 
  1.675 &   0.005 &   0.597 &   0.003 &     1.24 &     0.05 &      6.4   \\ 
  3.830 &   0.005 &   0.261 &   0.001 &     1.08 &     0.07 &     18.0   \\ 
  3.470 &   0.005 &   0.288 &   0.001 &     0.79 &     0.07 &     13.1   \\ 
  3.535 &   0.005 &   0.283 &   0.001 &     0.67 &     0.07 &     11.2   \\ 
  3.925 &   0.005 &   0.255 &   0.001 &     0.44 &     0.07 &      7.3   \\ 
  3.268 &   0.005 &   0.306 &   0.002 &     0.43 &     0.07 &      7.2   \\ 
  4.296 &   0.005 &   0.233 &   0.001 &     0.37 &     0.07 &      6.8   \\ 
\hline
\multicolumn{7}{c}{RX\,J0440.9+4431 (19)} \\
\hline
  2.599 &   0.010 &   0.385 &   0.004 &     8.12 &     0.21 &     16.8   \\ 
  2.174 &   0.010 &   0.460 &   0.005 &     6.46 &     0.21 &     13.4   \\ 
  1.098 &   0.010 &   0.911 &   0.009 &     3.14 &     0.21 &      6.5   \\ 
  6.738 &   0.010 &   0.148 &   0.002 &     1.56 &     0.45 &     20.4   \\ 
\hline
\multicolumn{7}{c}{1A\,0535+262 (43-45)} \\
\hline
  2.132 &   0.003 &   0.469 &   0.002 &     9.24 &     0.04 &     55.6   \\ 
  2.166 &   0.003 &   0.462 &   0.002 &     3.66 &     0.04 &     22.0   \\ 
  0.996 &   0.003 &   1.005 &   0.003 &     1.19 &     0.04 &      7.2   \\ 
  1.052 &   0.003 &   0.951 &   0.003 &     1.36 &     0.04 &      8.2   \\ 
  1.898 &   0.003 &   0.527 &   0.002 &     1.59 &     0.04 &      9.6   \\ 
  7.183 &   0.003 &   0.139 &   0.001 &     0.33 &     0.13 &      9.8   \\ 
  3.170 &   0.003 &   0.315 &   0.001 &     0.32 &     0.13 &      5.5   \\ 
  3.249 &   0.003 &   0.308 &   0.001 &     0.29 &     0.13 &      4.9   \\ 
  4.220 &   0.003 &   0.237 &   0.001 &     0.29 &     0.13 &      7.2   \\ 
  3.873 &   0.003 &   0.258 &   0.001 &     0.28 &     0.13 &      4.7   \\ 
  3.994 &   0.003 &   0.250 &   0.001 &     0.30 &     0.13 &      5.1   \\ 
  6.101 &   0.003 &   0.164 &   0.001 &     0.28 &     0.13 &      8.0   \\ 
  3.464 &   0.003 &   0.289 &   0.001 &     0.29 &     0.13 &      4.9   \\ 
  3.701 &   0.003 &   0.270 &   0.001 &     0.26 &     0.13 &      4.4   \\ 
\hline
\multicolumn{7}{c}{IGR\,J06074+2205 (43,44)} \\
\hline
  2.302 &   0.005 &   0.434 &   0.002 &     5.90 &     0.04 &     72.5   \\ 
  6.831 &   0.005 &   0.146 &   0.001 &     0.45 &     0.04 &      9.4   \\ 
  2.151 &   0.005 &   0.465 &   0.002 &     0.37 &     0.04 &      4.6   \\ 
  0.602 &   0.005 &   1.661 &   0.008 &     0.36 &     0.04 &      4.4   \\ 
  1.652 &   0.005 &   0.605 &   0.003 &     0.35 &     0.04 &      4.3   \\ 
%  4.607 &   0.005 &   0.217 &   0.001 &     0.34 &     0.09 &      7.5   \\ 
  5.347 &   0.005 &   0.187 &   0.001 &     0.21 &     0.08 &      4.2   \\ 
  3.012 &   0.005 &   0.332 &   0.002 &     0.23 &     0.09 &      5.5   \\ 
  6.088 &   0.005 &   0.164 &   0.001 &     0.22 &     0.08 &      5.3   \\ 
\hline
\multicolumn{7}{c}{MXB\,0656--072 (7)} \\
\hline
  1.383 &   0.010 &   0.723 &   0.007 &    12.80 &     0.06 &    103.7   \\ 
  2.764 &   0.010 &   0.362 &   0.004 &     1.96 &     0.05 &     15.9   \\ 
  0.170 &   0.010 &   5.881 &   0.060 &     1.11 &     0.06 &      9.0   \\ 
\hline
\multicolumn{7}{c}{RX\,J0812.4--3114 (7,8)} \\
\hline
  2.886 &   0.005 &   0.346 &   0.002 &     4.68 &     0.11 &     20.6   \\ 
  2.563 &   0.005 &   0.390 &   0.002 &     4.26 &     0.11 &     18.7   \\ 
  1.306 &   0.005 &   0.766 &   0.004 &     3.08 &     0.11 &     13.6   \\ 
  2.462 &   0.005 &   0.406 &   0.002 &     1.36 &     0.11 &      6.0   \\ 
  3.000 &   0.005 &   0.333 &   0.002 &     1.68 &     0.28 &     17.2   \\ 
  3.871 &   0.005 &   0.258 &   0.001 &     0.78 &     0.24 &      8.0   \\ 
  3.109 &   0.005 &   0.322 &   0.002 &     2.00 &     0.30 &     20.4   \\ 
  3.183 &   0.005 &   0.314 &   0.002 &     2.18 &     0.27 &     22.3   \\ 
\hline
\end{tabular}
\end{table*}
%-----------------------------------------------------------------------------------------

%-----------------------------------------------------------------------------------------
\begin{table*}
\caption{Frequencies and periods of BeXBS (cont.).} %The numbers besides the source name indicate the TESS sector.}
\label{freq-tess3}
\center
\begin{tabular}{ccccccc}
\hline
\hline
Frequency	&$\sigma(f)$	&Period	&$\sigma(P)$	&Amplitude	&$\sigma(A)$	&S/N	\\
(\cd)		&(\cd)		&(days)	&(days)		&(ppt)		&(ppt)		&	\\
\hline
\multicolumn{7}{c}{RX\,J0812.4--3114 (34,35)} \\
\hline
  2.883 &   0.005 &   0.347 &   0.002 &     3.09 &     0.06 &     24.3   \\ 
  0.539 &   0.005 &   1.854 &   0.009 &     0.76 &     0.06 &      6.0   \\ 
  9.005 &   0.005 &   0.111 &   0.001 &     0.67 &     0.06 &      8.9   \\ 
  0.378 &   0.005 &   2.647 &   0.013 &     0.60 &     0.06 &      4.7   \\ 
 11.890 &   0.005 &   0.084 &   0.000 &     0.62 &     0.08 &      8.8   \\ 
  3.030 &   0.005 &   0.330 &   0.002 &     0.51 &     0.08 &      7.2   \\ 
  3.475 &   0.005 &   0.288 &   0.001 &     0.30 &     0.08 &      4.3   \\ 
\hline
\multicolumn{7}{c}{GRO\,J1008-57 (9,10)} \\
\hline
  0.305 &   0.005 &   3.282 &   0.016 &     0.65 &     0.06 &      7.4   \\ 
  0.604 &   0.005 &   1.656 &   0.008 &     0.75 &     0.06 &      8.5   \\ 
  0.229 &   0.005 &   4.362 &   0.021 &     0.87 &     0.06 &      9.8   \\ 
  2.982 &   0.005 &   0.335 &   0.002 &     0.50 &     0.06 &      5.7   \\ 
 10.192 &   0.005 &   0.098 &   0.001 &     0.35 &     0.06 &      5.4   \\ 
\hline
\multicolumn{7}{c}{GRO\,J1008-57 (36,37)} \\
\hline
  2.983 &   0.005 &   0.335 &   0.002 &     0.42 &     0.05 &      5.7   \\ 
  1.334 &   0.005 &   0.750 &   0.004 &     0.34 &     0.05 &      4.7   \\ 
 10.195 &   0.005 &   0.098 &   0.001 &     0.31 &     0.05 &      5.5   \\ 
  1.166 &   0.005 &   0.858 &   0.004 &     0.30 &     0.05 &      4.1   \\ 
\hline
\multicolumn{7}{c}{ RX\,J1037.5--5647 (10)} \\
\hline
  0.319 &   0.009 &   3.132 &   0.030 &     3.18 &     0.18 &      7.4   \\ 
  0.943 &   0.009 &   1.060 &   0.010 &     3.21 &     0.18 &      7.4   \\ 
  1.103 &   0.009 &   0.906 &   0.009 &     2.14 &     0.18 &      5.0   \\ 
  4.912 &   0.009 &   0.204 &   0.002 &     0.92 &     0.25 &      8.3   \\ 
  3.775 &   0.009 &   0.265 &   0.003 &     0.87 &     0.25 &      8.5   \\ 
  3.877 &   0.009 &   0.258 &   0.003 &     0.77 &     0.25 &      7.5   \\ 
  5.649 &   0.009 &   0.177 &   0.002 &     0.53 &     0.25 &      6.0   \\ 
\hline
\multicolumn{7}{c}{ RX\,J1037.5--5647 (36,37)} \\
\hline
  0.305 &   0.005 &   3.282 &   0.016 &     2.36 &     0.09 &      6.7   \\ 
  1.863 &   0.005 &   0.537 &   0.003 &     2.38 &     0.09 &      6.8   \\ 
  0.454 &   0.005 &   2.204 &   0.011 &     2.15 &     0.09 &      6.1   \\ 
  0.793 &   0.005 &   1.261 &   0.006 &     2.05 &     0.09 &      5.9   \\ 
  4.914 &   0.005 &   0.203 &   0.001 &     1.10 &     0.12 &     13.5   \\ 
%  3.722 &   0.005 &   0.269 &   0.001 &     0.82 &     0.12 &      8.3   \\ 
  3.095 &   0.005 &   0.323 &   0.002 &     0.54 &     0.12 &      5.4   \\ 
  3.281 &   0.005 &   0.305 &   0.002 &     0.51 &     0.12 &      5.2   \\ 
  3.380 &   0.005 &   0.296 &   0.001 &     0.44 &     0.12 &      4.5   \\ 
  5.638 &   0.005 &   0.177 &   0.001 &     0.41 &     0.12 &      6.8   \\ 
  3.200 &   0.005 &   0.312 &   0.002 &     0.50 &     0.12 &      5.0   \\ 
\hline
\multicolumn{7}{c}{1A\,1118--616 (10,11)} \\
\hline
  0.934 &   0.005 &   1.071 &   0.005 &     1.40 &     0.05 &     12.6   \\ 
  1.163 &   0.005 &   0.860 &   0.004 &     1.34 &     0.05 &     12.1   \\ 
  0.299 &   0.005 &   3.348 &   0.015 &     1.08 &     0.05 &      9.7   \\ 
  6.531 &   0.005 &   0.153 &   0.001 &     0.38 &     0.07 &     12.8   \\ 
  4.423 &   0.005 &   0.226 &   0.001 &     0.33 &     0.07 &     10.2   \\ 
  3.248 &   0.005 &   0.308 &   0.001 &     0.33 &     0.07 &      9.4   \\ 
  3.664 &   0.005 &   0.273 &   0.001 &     0.21 &     0.07 &      6.1   \\ 
  3.102 &   0.005 &   0.322 &   0.002 &     0.22 &     0.07 &      6.3   \\ 
\hline
\multicolumn{7}{c}{1A\,1118--616 (37,38)} \\
\hline
  0.326 &   0.005 &   3.068 &   0.014 &     4.35 &     0.03 &     48.3   \\ 
  0.934 &   0.005 &   1.071 &   0.005 &     0.93 &     0.03 &     10.3   \\ 
  0.398 &   0.005 &   2.512 &   0.012 &     0.59 &     0.03 &      6.6   \\ 
 14.015 &   0.005 &   0.071 &   0.000 &     0.22 &     0.06 &      8.9   \\ 
  3.128 &   0.005 &   0.320 &   0.002 &     0.19 &     0.07 &      6.2   \\ 
  6.530 &   0.005 &   0.153 &   0.001 &     0.19 &     0.06 &      8.0   \\ 
  4.422 &   0.005 &   0.226 &   0.001 &     0.17 &     0.06 &      6.9   \\ 
  3.244 &   0.005 &   0.308 &   0.001 &     0.13 &     0.06 &      4.2   \\ 
  3.050 &   0.005 &   0.328 &   0.002 &     0.15 &     0.06 &      5.1   \\ 
\hline
\end{tabular}
\end{table*}
%-----------------------------------------------------------------------------------------

%-----------------------------------------------------------------------------------------
\begin{table*}
\caption{Frequencies and periods of BeXBS (cont.). } 
\label{freq-tess4}
\center
\begin{tabular}{ccccccc}
\hline
\hline
Frequency	&$\sigma(f)$	&Period	&$\sigma(P)$	&Amplitude	&$\sigma(A)$	&S/N	\\
(\cd)		&(\cd)		&(days)	&(days)		&(ppt)		&(ppt)		&	\\
\hline
\multicolumn{7}{c}{4U\,1145--61 (10,11)} \\
\hline
  2.155 &   0.005 &   0.464 &   0.002 &     1.48 &     0.08 &      8.7   \\ 
  1.502 &   0.005 &   0.666 &   0.003 &     1.28 &     0.08 &      7.5   \\ 
  2.103 &   0.005 &   0.475 &   0.002 &     1.02 &     0.08 &      6.0   \\ 
  0.383 &   0.005 &   2.610 &   0.012 &     0.98 &     0.08 &      5.8   \\ 
  3.217 &   0.005 &   0.311 &   0.001 &     0.84 &     0.10 &     11.6   \\ 
  3.063 &   0.005 &   0.327 &   0.002 &     0.71 &     0.10 &      9.9   \\ 
  3.158 &   0.005 &   0.317 &   0.002 &     0.68 &     0.10 &      9.4   \\ 
  4.566 &   0.005 &   0.219 &   0.001 &     0.42 &     0.09 &      7.9   \\ 
  8.652 &   0.005 &   0.116 &   0.001 &     0.42 &     0.09 &      9.9   \\ 
  3.625 &   0.005 &   0.276 &   0.001 &     0.37 &     0.09 &      5.1   \\ 
  8.238 &   0.005 &   0.121 &   0.001 &     0.30 &     0.09 &      7.0   \\ 
  4.379 &   0.005 &   0.228 &   0.001 &     0.26 &     0.09 &      4.9   \\ 
 \hline
\multicolumn{7}{c}{4U\,1145--61 (37,38)} \\
\hline
  2.795 &   0.005 &   0.358 &   0.002 &     2.55 &     0.04 &     19.4   \\ 
  2.160 &   0.005 &   0.463 &   0.002 &     2.47 &     0.04 &     18.8   \\ 
  2.104 &   0.005 &   0.475 &   0.002 &     1.52 &     0.04 &     11.6   \\ 
  3.244 &   0.005 &   0.308 &   0.001 &     0.70 &     0.06 &     12.0   \\ 
  4.560 &   0.005 &   0.219 &   0.001 &     0.55 &     0.06 &     10.0   \\ 
  3.024 &   0.005 &   0.331 &   0.002 &     0.55 &     0.06 &      9.4   \\ 
  8.651 &   0.005 &   0.116 &   0.001 &     0.47 &     0.06 &     12.2   \\ 
 12.698 &   0.005 &   0.079 &   0.000 &     0.32 &     0.06 &     11.4   \\ 
 10.221 &   0.005 &   0.098 &   0.001 &     0.30 &     0.06 &      8.0   \\ 
  4.270 &   0.005 &   0.234 &   0.001 &     0.28 &     0.06 &      5.1   \\ 
  3.316 &   0.005 &   0.302 &   0.001 &     0.29 &     0.06 &      5.0   \\ 
  8.909 &   0.005 &   0.112 &   0.001 &     0.25 &     0.06 &      6.5   \\ 
 10.395 &   0.005 &   0.096 &   0.001 &     0.24 &     0.06 &      6.5   \\ 
\hline
\multicolumn{7}{c}{GX\,304--1 (38)} \\
\hline
  1.802 &   0.009 &   0.555 &   0.005 &     6.17 &     0.12 &     12.9   \\ 
  6.768 &   0.009 &   0.148 &   0.001 &     3.24 &     0.12 &     21.4   \\ 
  7.185 &   0.009 &   0.139 &   0.001 &     3.13 &     0.12 &     20.7   \\ 
  1.018 &   0.009 &   0.983 &   0.009 &     2.14 &     0.12 &      4.5   \\ 
  3.018 &   0.009 &   0.331 &   0.003 &     1.79 &     0.16 &     14.1   \\ 
  3.601 &   0.009 &   0.278 &   0.003 &     0.96 &     0.16 &      7.6   \\ 
  8.559 &   0.009 &   0.117 &   0.001 &     0.88 &     0.16 &      9.3   \\ 
\hline
\multicolumn{7}{c}{KS\,1947+300 (41)} \\
\hline
  0.498 &   0.009 &   2.006 &   0.019 &     0.92 &     0.12 &      5.0   \\ 
  3.652 &   0.009 &   0.274 &   0.003 &     0.68 &     0.12 &      6.2   \\ 
\hline
\multicolumn{7}{c}{GRO\,J2058+42 (15)} \\
\hline
  2.373 &   0.011 &   0.421 &   0.005 &     7.36 &     0.29 &     13.3   \\ 
  0.290 &   0.011 &   3.450 &   0.039 &     3.83 &     0.29 &      6.9   \\ 
  6.375 &   0.011 &   0.157 &   0.002 &     2.69 &     0.29 &     10.2   \\ 
  3.994 &   0.011 &   0.250 &   0.003 &     1.63 &     0.42 &      6.8   \\ 
\hline
\multicolumn{7}{c}{4U\,2206+54 (16,17)} \\
\hline
  0.590 &   0.005 &   1.696 &   0.008 &     3.35 &     0.13 &     11.2   \\ 
  1.008 &   0.005 &   0.992 &   0.005 &     2.49 &     0.13 &      8.3   \\ 
  0.279 &   0.005 &   3.585 &   0.018 &     2.16 &     0.13 &      7.2   \\ 
  0.511 &   0.005 &   1.956 &   0.010 &     1.84 &     0.13 &      6.1   \\ 
  0.810 &   0.005 &   1.234 &   0.006 &     2.23 &     0.13 &      7.4   \\ 
  1.264 &   0.005 &   0.791 &   0.004 &     1.76 &     0.13 &      5.8   \\ 
  6.136 &   0.005 &   0.163 &   0.001 &     0.85 &     0.20 &     14.2   \\ 
  3.284 &   0.005 &   0.304 &   0.002 &     0.75 &     0.20 &      7.9   \\ 
  3.067 &   0.005 &   0.326 &   0.002 &     0.66 &     0.20 &      7.0   \\ 
  3.388 &   0.005 &   0.295 &   0.001 &     0.60 &     0.20 &      6.4   \\ 
  3.121 &   0.005 &   0.320 &   0.002 &     0.51 &     0.20 &      5.4   \\ 
  3.682 &   0.005 &   0.272 &   0.001 &     0.49 &     0.20 &      5.1   \\ 
  3.963 &   0.005 &   0.252 &   0.001 &     0.44 &     0.20 &      4.7   \\ 
\hline
\end{tabular}
\end{table*}
%-----------------------------------------------------------------------------------------

%-----------------------------------------------------------------------------------------
\begin{table*}[h]
\caption{Frequencies and periods of BeXBs (cont.).} 
\label{freq-tess5}
\center
\begin{tabular}{ccccccc}
\hline
\hline
Frequency	&$\sigma(f)$	&Period	&$\sigma(P)$	&Amplitude	&$\sigma(A)$	&S/N	\\
(\cd)		&(\cd)		&(days)	&(days)		&(ppt)		&(ppt)		&	\\
\hline
\multicolumn{7}{c}{SAX\,J2239.3+6116 (16,17)} \\
\hline
  0.731 &   0.005 &   1.369 &   0.007 &     5.85 &     0.06 &     59.6   \\ 
  0.303 &   0.005 &   3.297 &   0.016 &     1.76 &     0.06 &     17.9   \\ 
  0.663 &   0.005 &   1.508 &   0.007 &     1.79 &     0.06 &     18.2   \\ 
  2.011 &   0.005 &   0.497 &   0.002 &     0.46 &     0.16 &      5.6   \\ 
  2.082 &   0.005 &   0.480 &   0.002 &     0.39 &     0.16 &      4.7   \\ 
\hline
\multicolumn{7}{c}{SAX\,J2239.3+6116 (24)} \\
\hline
  0.163 &   0.009 &   6.146 &   0.058 &     3.61 &     0.13 &     17.5   \\ 
  0.727 &   0.009 &   1.376 &   0.013 &     2.51 &     0.13 &     12.2   \\ 
  1.934 &   0.009 &   0.517 &   0.002 &     1.60 &     0.13 &      7.8   \\
% 10.346 &   0.009 &   0.097 &   0.001 &     0.85 &     0.24 &      9.7   \\ 
 11.164 &   0.009 &   0.090 &   0.001 &     0.53 &     0.23 &      4.8   \\ 
\hline
\multicolumn{7}{c}{MWC\,656 (16)} \\
\hline
  1.791 &   0.010 &   0.558 &   0.006 &     4.46 &     0.15 &     13.1   \\ 
  0.775 &   0.010 &   1.291 &   0.013 &     3.14 &     0.16 &      9.2   \\ 
  1.087 &   0.010 &   0.920 &   0.009 &     1.59 &     0.15 &      4.7   \\ 
  3.690 &   0.010 &   0.271 &   0.003 &     1.32 &     0.26 &     29.5   \\ 
  4.301 &   0.010 &   0.233 &   0.002 &     0.68 &     0.26 &     17.0   \\ 
\hline
\end{tabular}
\end{table*}
%-----------------------------------------------------------------------------------------

%-----------------------------------------------------------------------------------------
\begin{table*}[h]
\caption{Frequencies and periods of SGXBs. The numbers besides the source name indicate the TESS sector.}
\label{freq-sgxb}
\center
\begin{tabular}{ccccccc}
\hline
\hline
Frequency	&$\sigma(f)$	&Period	&$\sigma(P)$	&Amplitude	&$\sigma(A)$	&S/N	\\
(\cd)		&(\cd)		&(days)	&(days)		&(ppt)		&(ppt)		&	\\
\hline
\multicolumn{7}{c}{IGR\,J00370+6122 (17,18)} \\
\hline
  0.203 &   0.005 &   4.919 &   0.024 &     6.49 &     0.10 &     31.8   \\ 
  0.407 &   0.005 &   2.458 &   0.012 &     4.20 &     0.09 &     20.6   \\ 
  0.556 &   0.005 &   1.799 &   0.009 &     3.99 &     0.09 &     19.6   \\ 
  0.126 &   0.005 &   7.951 &   0.040 &     3.00 &     0.09 &     14.7   \\ 
\hline
\multicolumn{7}{c}{IGR\,J00370+6122 (24)} \\
\hline
  0.363 &   0.009 &   2.751 &   0.026 &     4.84 &     0.15 &     14.3   \\ 
  0.713 &   0.009 &   1.403 &   0.013 &     3.56 &     0.15 &     10.5   \\ 
  0.158 &   0.009 &   6.344 &   0.060 &     4.31 &     0.16 &     12.7   \\ 
  3.064 &   0.009 &   0.326 &   0.003 &     0.69 &     0.36 &      8.7   \\ 
\hline
\multicolumn{7}{c}{2S\,0114+65 (18)} \\
\hline
  0.149 &   0.010 &   6.700 &   0.069 &    20.52 &     0.08 &    154.0   \\ 
  0.349 &   0.010 &   2.864 &   0.029 &     6.27 &     0.07 &     47.0   \\ 
\hline
\multicolumn{7}{c}{2S\,0114+65 (24,25)} \\
\hline
  0.151 &   0.005 &   6.622 &   0.031 &    10.96 &     0.05 &    117.5   \\ 
  0.398 &   0.005 &   2.510 &   0.012 &     5.35 &     0.05 &     57.4   \\ 
\hline
\multicolumn{7}{c}{Vela X--1 (8,9)} \\
\hline
  0.222 &   0.005 &   4.498 &   0.022 &    31.70 &     0.03 &    882.8   \\ 
  0.117 &   0.005 &   8.512 &   0.042 &     9.86 &     0.04 &    274.6   \\ 
  0.038 &   0.005 &  26.288 &   0.129 &     6.22 &     0.03 &    173.2   \\ 
  0.335 &   0.005 &   2.981 &   0.015 &    10.12 &     0.03 &    281.7   \\ 
  0.750 &   0.005 &   1.334 &   0.006 &     5.42 &     0.02 &    151.0   \\ 
\hline
\end{tabular}
\end{table*}
%-----------------------------------------------------------------------------------------

\section{Log of the ground-based observations}
\label{ground}

This appendix summarizes the ground-based observations. We give the date of the observations, as modified Julian date and civil date, the number of data points per night and the filter used.

%-----------------------------------------------------------------------------------------
\begin{table}
\caption{Log of observations. }
\label{logobs1}
\center
\begin{tabular}{lllc}
\hline
\hline
MJD	&Date		& Site	&$N_{\rm obs}$ (filter)  \\
\hline
\multicolumn{4}{c}{4U\,0115+63} \\
\hline
54725	&15-09-2008	&SKO	&45 (V) \\
54726	&16-09-2008	&SKO	&49 (V) \\
54755	&15-10-2008	&SKO	&52 (V) \\
54756	&16-10-2008	&SKO	&84 (V) \\
55097	&22-09-2009	&SKO	&21 (V) \\
55098	&23-09-2009	&SKO	&49 (V) \\
55099	&24-09-2009	&SKO	&26 (V) \\
55100	&25-09-2009	&SKO	&3  (V)  \\
55101	&26-09-2009	&SKO	&24 (V) \\
55102	&27-09-2009	&SKO	&33 (V) \\
55109	&04-10-2009	&SKO	&28 (V) \\
55110	&05-10-2009	&SKO	&26 (V) \\
55112	&07-10-2009	&SKO	&8  (V)  \\
55113	&08-10-2009	&SKO	&35 (V) \\
55114	&09-10-2009	&SKO	&41 (V) \\
55115	&10-10-2009	&SKO	&44 (V) \\
55116	&11-10-2009	&SKO	&44 (V) \\
\hline
 \multicolumn{4}{c}{RX J0146.9+6121/LS I +61 235} \\
\hline
54725	&15-09-2008	&SKO	&78 (V)       \\
54726	&16-09-2008	&SKO	&75 (V)       \\
58382	&20-09-2018	&SKO	&545 (V), 546 (B) \\
58383	&21-09-2018	&SKO	&632 (V), 635 (B) \\
58384	&22-09-2018	&SKO	&631 (V), 493 (B) \\
58385	&23-09-2018	&SKO	&635 (V), 635 (B) \\
\hline
 \multicolumn{4}{c}{SWIFT\,J0243.6+6124} \\
\hline
58760	&03-10-2019	&AOO	&43 (V), 43 (B)  \\
58765	&08-10-2019	&AOO	&29 (V), 29 (B)  \\
58781	&24-10-2019	&AOO	&29 (V), 29 (B)  \\
58782	&25-10-2019	&AOO	&45 (V), 46 (B)  \\
58783	&26-10-2019	&AOO	&37 (V), 38 (B)  \\
\hline
 \multicolumn{4}{c}{RX J0441.0+4431/LS\,V 4417} \\
\hline
58494	&10-01-2019	&AOO	&7 (V), 75 (R)  \\
58495	&11-01-2019	&AOO	&316 (V), 316 (R) \\
58496	&12-01-2019	&AOO	&354 (V), 353 (R) \\
58498	&14-01-2019	&AOO	&453 V), 453 (R) \\
\hline
 \multicolumn{4}{c}{1A\,0535+262} \\
\hline
55109	&04-10-2009	&SKO	&116 (V) \\
55110	&05-10-2009	&SKO	&58  (V) \\
55113	&08-10-2009	&SKO	&69  (V) \\
55114	&09-10-2009	&SKO	&155 (V) \\
55115	&10-10-2009	&SKO	&161 (V) \\
55116	&11-10-2009	&SKO	&152 (V) \\
\hline
 \multicolumn{4}{c}{IGR\,J06074+2205} \\
\hline
58520	&05-02-2019	&AOO	&40 (V), 38 (R)  \\
58521	&06-02-2019	&AOO	&37 (V), 35 (R)  \\
58523	&08-02-2019	&AOO	&46 (V), 44 (R)  \\
58526	&11-02-2019	&AOO	&99 (V), 97 (R)  \\
58527	&12-02-2019	&AOO	&66 (V), 65 (R)  \\
\hline
 \multicolumn{4}{c}{KS\,1947+300} \\
\hline
55406	&28-07-2010	&SKO	&67 (V)	\\
55407	&29-07-2010	&SKO	&74 (V)	\\
55408	&30-07-2010	&SKO	&67 (V)	\\
55778	&04-08-2011	&SKO	&90 (V)	\\
55779	&05-08-2011	&SKO	&94 (V)	\\
55780	&06-08-2011	&SKO	&45 (V)	\\
\hline
\end{tabular}
\end{table}
%-----------------------------------------------------------------------------------------

%-----------------------------------------------------------------------------------------
\begin{table}
\caption{Log of observations (cont). }
\label{logobs2}
\center
\begin{tabular}{lllc}
\hline
\hline
MJD	&Date		& Site	&$N_{\rm obs}$ (filter)  \\
\hline
 \multicolumn{4}{c}{Swift J2000.6+3210} \\
\hline
59424	&28-07-2021	&SKO	&107 (V) \\
59425	&29-07-2021	&SKO	&78 (V) \\
59426	&30-07-2021	&SKO	&37 (V) \\
59427	&31-07-2021	&SKO	&72 (V) \\
59446	&19-08-2021	&SKO	&104 (V) \\
59447	&20-08-2021	&SKO	&112 (V) \\
59448	&21-08-2021	&SKO	&111 (V) \\
59449	&22-08-2021	&SKO	&48 (V) \\
\hline
 \multicolumn{4}{c}{GRO\,J2058+42} \\
\hline
59046	&15-07-2020	&SKO	&149 (V)\\
59047	&16-07-2020	&SKO	&150 (V)\\
59048	&17-07-2020	&SKO	&117 (V)\\
59049	&18-07-2020	&SKO	&145 (V)\\
\hline
 \multicolumn{4}{c}{SAX\,J2103.5+4545} \\
\hline
55097	&22-09-2009	&SKO	&18 (V) \\
55098	&23-09-2009	&SKO	&27 (V) \\
55099	&24-09-2009	&SKO	&31 (V) \\
55100	&25-09-2009	&SKO	&3 (V)  \\
55101	&26-09-2009	&SKO	&25 (V) \\
55102	&27-09-2009	&SKO	&27 (V) \\
55109	&04-10-2009	&SKO	&13 (V) \\
55110	&05-10-2009	&SKO	&21 (V) \\
55112	&07-10-2009	&SKO	&6 (V)  \\
55113	&08-10-2009	&SKO	&27 (V) \\
55114	&09-10-2009	&SKO	&21 (V) \\
55115	&10-10-2009	&SKO	&22 (V) \\
55116	&11-10-2009	&SKO	&21 (V) \\
59058	&27-07-2020	&SKO	&293 (V) \\
59059	&28-07-2020	&SKO	&320 (V) \\
59060	&29-07-2020	&SKO	&400 (V) \\
59061	&30-07-2020	&SKO	&400 (V) \\
\hline
 \multicolumn{4}{c}{IGR\,J21343+4738} \\
\hline
58309	&09-07-2018	&AOO	&26 (V), 25 (R)  \\
58310	&10-07-2018	&AOO	&27 (V), 27 (R)  \\
58310	&10-07-2018	&SKO	&25 (V), 25 (R)    \\
58311	&11-07-2018	&SKO	&92 (V), 92 (R)    \\
58312	&12-07-2018	&SKO	&67 (V), 67 (R)    \\
58313	&13-07-2018	&AOO	&24 (V), 22 (R)  \\
58318	&18-07-2018	&AOO	&35 (V), 33 (R) \\
58696	&31-07-2019	&SKO	&61 (V), 61 (B) \\
58697	&01-08-2019	&SKO	&48 (V), 48 (B) \\
58698	&02-08-2019	&SKO	&63 (V), 63 (B) \\
58713	&17-08-2019	&AOO	&36 (V), 37 (B) \\
58717	&21-08-2019	&AOO	&43 (V), 44 (B) \\
58719	&23-08-2019	&AOO	&44 (V), 43 (B) \\
58721	&25-08-2019	&AOO	&28 (V), 28 (B) \\
58725	&29-08-2019	&AOO	&32 (V), 32 (B) \\
58727	&31-08-2019	&AOO	&51 (V), 51 (B) \\
58745	&18-09-2019	&AOO	&25 (V), 26 (B) \\
58749	&22-09-2019	&AOO	&34 (V), 34 (B) \\
58752	&25-09-2019	&AOO	&33 (V), 33 (B)  \\
\hline
\end{tabular}
\end{table}
%-----------------------------------------------------------------------------------------

%-----------------------------------------------------------------------------------------
\begin{table}
\caption{Log of observations (cont). }
\label{logobs3}
\center
\begin{tabular}{lllc}
\hline
\hline
MJD	&Date		& Site	&$N_{\rm obs}$ (filter)  \\
\hline
 \multicolumn{4}{c}{Cep\, X--4} \\
\hline
59084	&22-10-2020	&SKO	&240 (V)      \\
59085	&23-10-2020	&SKO	&250 (V)       \\
59086	&24-09-2020	&SKO	&237  (V)      \\
\hline
 \multicolumn{4}{c}{4U\,2206+54} \\
\hline
54755	&15-10-2008	&SKO	&138 (V)      \\
54756	&16-10-2008	&SKO	&177 (V)       \\
55097	&22-09-2009	&SKO	&65  (V)      \\
55098	&23-09-2009	&SKO	&106 (V)      \\
55099	&24-09-2009	&SKO	&87  (V)      \\
55100	&25-09-2009	&SKO	&9   (V)      \\
55101	&26-09-2009	&SKO	&87  (V)      \\
55102	&27-09-2009	&SKO	&104 (V)      \\
55109	&04-10-2009	&SKO	&72  (V)      \\
55110	&05-10-2009	&SKO	&100 (V)      \\
55112	&07-10-2009	&SKO	&43  (V)      \\
55113	&08-10-2009	&SKO	&103(V)       \\
55114	&09-10-2009	&SKO	&79 (V)       \\
55115	&10-10-2009	&SKO	&89 (V)       \\
55116	&11-10-2009	&SKO	&84 (V)       \\
\hline
 \multicolumn{4}{c}{SAX\,J2239.3+6116} \\
\hline
55796	&22-08-2011	&SKO	&100 (V) \\
55797	&23-08-2011	&SKO	&100 (V)\\
55798	&24-08-2011	&SKO	&100 (V)\\
55799	&25-08-2011	&SKO	&100 (V) \\
58716	&20-08-2019	&SKO	&40 (V), 40 (B) \\
58717	&21-08-2019	&SKO	&40 (V), 40 (B) \\
58718	&22-08-2019	&SKO	&41 (V), 41 (B) \\
58719	&23-08-2019	&SKO	&35 (V), 35 (B) \\
\hline
 \multicolumn{4}{c}{MWC\,656} \\
\hline
58758	&01-10-2019	&AOO	&264 (V), 280 (B)  \\
58759	&02-10-2019	&AOO	&228 (V), 244 (B)  \\
58761	&04-10-2019	&AOO	&269 (V), 267 (B)  \\
58763	&06-10-2019	&AOO	&190 (V), 192 (B)  \\
58764	&07-10-2019	&AOO	&227 (V), 204 (B)  \\
\hline
\end{tabular}
\end{table}
%-----------------------------------------------------------------------------------------

\section{Detected frequencies from ground-based observations}
\label{ground-freq}

Table~\ref{freq-ski} gives the significant frequencies detected in the data obtained from the Skinakas observatory.

%-----------------------------------------------------------------------------------------
\begin{table}
\caption{Detected periodicities in Be/X-ray binaries from ground-based photometry.}
\label{freq-ski}
\center
\begin{tabular}{@{~~}l@{~~}c@{~~}c@{~~}c@{~~}c}
\hline
\hline
System			&Frequency	&Period	&Amplitude	&S/N	\\
			&(\cd)		&(hour)	&(mmag)		&     \\
\hline		
4U\,0115+63 (2008)	&3.40		&7.06	&17.0		&8.1  \\
4U\,0115+63 (2009)	&3.33		&7.21	&13.4		&8.5  \\
RX\,J0146.9+6121	&2.90		&8.28	&6.1		&6.2  \\
			&9.70		&2.47	&4.1		&8.3  \\
			&1.36		&17.6   &3.7		&3.7  \\
Swift\,J0243.6+6124	&--		&--	&--		&--   \\
RX\,J0440.9+4431	&3.20		&7.50	&10		&3.5 \\
1A\,0535+262	&--		&--	&--		&--  \\
IGR\,J06074+2205	&--		&--	&--		&--  \\
%			&0.51		&	&18.7		& \\
%KS\,1947+300 (2010)	&--		&--	&--		&--   \\
KS\,1947+300 (2011)	&2.31		&10.4	&3.9		&12.5   \\
			&2.73		&8.79	&2.2		&6.5    \\
%			&8.40		&	&0.7		&4.7  \\
%EXO\,2030+375	&		&	&		&     \\
Swift\,J2000.6+3210	&2.79		&8.60	&6.6		&10 \\
%			&34.7		&	&1.9		&4.4 \\
GRO\,J2058+42 (2020)	&2.40		&10.0	&5.3		&13.0  \\
			&2.64		&9.09	&5.0		&12.0 \\
			&3.97		&6.05	&3.0		&6.6 \\
SAX\,J2103.5+4545 (2009)&3.22		&7.45	&4.1		&4.9  \\
SAX\,J2103.5+4545 (2020)&2.70		&8.89	&2.3		&8.3  \\
			&2.40		&10.0	&1.4		&5.6 \\
IGR\,J21343+4738 (2018)	&3.66		&6.56	&5.0		&4.0  \\
%			&2.04		&11.8	&3.1		&4.5  \\
%IGR\,J21343+4738--2019-S&2.74		&	&4.3		&5.5   \\
%IGR\,J21343+4738--2019-A&2.77		&	&6.5		&5.2   \\
IGR\,J21343+4738 (2019) &2.74		&8.76	&5.2		&5.7   \\
			&4.69		&5.12	&3.9		&4.7 \\
%			&2.08		&11.5	&2.8		&4.9 \\
CEP X--4		&2.07		&11.6	&3.6		&8.0  \\
			&5.37		&4.47	&2.4		&5.1  \\
4U\,2206+54 (2008)	&2.63		&9.13	&5.3		&30.0  \\
			&4.86		&4.94	&3.2		&8.2   \\
4U\,2206+54 (2009)	&2.12		&11.3	&3.6		&5.8  \\
			&2.60		&9.23	&5.3		&3.9  \\
SAX\,J2239.3+6116 (2011)&1.98		&12.1	&9.6		&17.4 \\
SAX\,J2239.3+6116 (2019)&--		&--	&--		&-- \\
MWC\,656		&2.78		&8.63	&9.3		&7.5  \\
			&1.97		&12.2	&7.9		&7.5  \\
\hline
\end{tabular}
\end{table}
%-----------------------------------------------------------------------------------------

\end{appendix}

\end{document}